\documentclass[aps,pra,epsfigure,twocolumn]{revtex4-2}
\usepackage[colorlinks=true,linkcolor=blue,urlcolor=blue,citecolor=blue,pdfusetitle]{hyperref}
\usepackage[utf8]{inputenc}
\usepackage[english]{babel}
\usepackage{amsmath}
\usepackage[caption = false]{subfig}
\usepackage{graphicx,epstopdf}
\usepackage{blindtext}
\usepackage[table,xcdraw]{xcolor}
\usepackage{lipsum}
\usepackage{amsfonts}
\usepackage{bbm}
\usepackage{amssymb}
\usepackage{enumerate}
\usepackage{color}
\usepackage{latexsym}
\usepackage{physics}
\usepackage{times,txfonts}
\usepackage{soul}

\newcommand{\Ccal}{\mathcal{C}}

\newcommand{\Ecal}{\mathcal{E}}

\newcommand{\1}{\mathbbm{1}}

\newcommand{\Zmath}{\mathbbm{Z}}

\usepackage{orcidlink}

\begin{document}

\title{Localization effects in disordered quantum batteries}

\author{Mohammad B. Arjmandi~\!\!\orcidlink{0000-0002-9222-6765}}
\email{m.arjmandi@sci.ui.ac.ir} 
\affiliation{Faculty of Physics, University of Isfahan, P.O. Box 81746-7344, Isfahan, Iran.}
\affiliation{Quantum Optics Research Group, University of Isfahan, Isfahan, Iran}

\author{Hamidreza Mohammadi~\!\!\orcidlink{0000-0001-7046-3818}}
\email{hr.mohammadi@sci.ui.ac.ir}
\affiliation{Faculty of Physics, University of Isfahan, P.O. Box 81746-7344, Isfahan, Iran.}
\affiliation{Quantum Optics Research Group, University of Isfahan, Isfahan, Iran}

\author{Andreia Saguia~\!\!\orcidlink{0000-0003-0403-4358}}
\email{asaguia@id.uff.br}
\affiliation{Instituto de F\'{i}sica, Universidade Federal Fluminense, Av. Gal. Milton Tavares de Souza s/n, Gragoat\'{a}, 24210-346 Niter\'{o}i, Rio de Janeiro, Brazil}

\author{Marcelo S. Sarandy~\!\!\orcidlink{0000-0003-0910-4407}}
\email{msarandy@id.uff.br}
\affiliation{Instituto de F\'{i}sica, Universidade Federal Fluminense, Av. Gal. Milton Tavares de Souza s/n, Gragoat\'{a}, 24210-346 Niter\'{o}i, Rio de Janeiro, Brazil}

\author{Alan C. Santos~\!\!\orcidlink{0000-0002-6989-7958}}
\email{ac\_santos@df.ufscar.br}
\affiliation{Departamento de Física, Universidade Federal de São Carlos, Rodovia Washington Luís, km 235 - SP-310, 13565-905 São Carlos, SP, Brazil}


\begin{abstract}
We investigate the effect of localization on the local charging of quantum batteries (QBs) modeled by disordered spin systems. 
Two distinct schemes based on the transverse-field random Ising model are considered, with Ising couplings defined on a Chimera graph and on a linear chain with up to next-to-nearest neighbor interactions. 
By adopting a low-energy demanding charging process driven by local fields only, we obtain that the maximum extractable energy by unitary processes (ergotropy) is highly enhanced in the ergodic phase in 
comparison with the many-body localization (MBL) scenario. As we turn off the next-to-nearest neighbor interactions in the Ising chain, we have the onset of the Anderson localization phase. We then show that the Anderson phase 
exhibits a hybrid behavior, interpolating between large and small ergotropy as the disorder strength is increased. We also consider the splitting of total ergotropy into its 
coherent and incoherent contributions. This incoherent part implies in a residual ergotropy that is fully robust against dephasing, which is a typical process leading to the self-discharging of the battery in a real setup. 
Our results are experimentally feasible in scalable systems, such as in superconducting integrated circuits.
\end{abstract}

\maketitle

\section{Introduction}

Few-body devices able to exploit quantum features, such as entanglement and coherence, have been referred to as quantum devices, leading to the second revolution for quantum technologies. 
They can potentially provide striking advantages, such as more sensitivity~\cite{degen2017quantum,zhang2021distributed,crawford2021quantum,feng2019review} and 
high efficient metrology~\cite{PhysRevLett.128.240401,giovannetti2011advances,PhysRevLett.96.010401}. They can also be used for communication tasks~\cite{Bennett:93,yin2012quantum,ma2012quantum,Ren:17}, 
heat and spin current devices like transistors~\cite{geppert2000quantum,Marchukov:16,joulain2016quantum,Loft:18,dePonte:19} and diodes~\cite{mcrae2019graphene,shukla2008nonlinear,shen2014quantum}, 
among others~\cite{Devoret:98,Arute:19,Bacon:09,Santos:16,Gottesman:99,PhysRevLett.81.5932,divincenzo1999quantum}. Naturally, each of these devices needs energy to operate and this energy, in turn, 
can be supplied by another quantum storage device, i.e., a quantum battery (QB)~\cite{campaioli2018quantum,bhattacharjee2021quantum}. By following the seminal work by Alicki and Fannes~\cite{Alicki:13}, 
QBs are designed as a set of interacting quantum cells, each of them usually represented by a two-level quantum system (qubit). These cells can be charged to store energy (with respect to a reference Hamiltonian) 
by means of external agents, such as fields, interactions, or any other stimulus able to inject energy into the system. Although QBs are typically composed by qubits, proposals of QB with qutrits~\cite{Santos:19-a} 
and bosonic systems can also be found~\cite{Andolina:19,Shaghaghi:micromaser,Downing_2022}. 

QBs may exhibit supercharging behavior, with efficiency of energy extraction growing with the number of cells~\cite{Alicki:13}. Such advantage with respect to classical batteries has been attributed 
to entanglement and quantum correlations~\cite{PRL_Andolina,PRB2019Batteries,arjmandi2022performance,PRL2017Binder,Binder:15,Rossini:20,Crescente:20,Ferraro:18}. 
However, both constructive and destructive scenarios have been found for the interplay between entanglement and available energy in QBs~\cite{Gianluca:22,Gianluca:17,PRL_Andolina,James:20,Kamin:20-2,Ghosh:21}. 
On the other hand, coherence seems to play an important role for the QB performance~\cite{Alexia:20,Francica:20,Kamin:20-2,Santos:21b,Baris:20,shi2022entanglement,2023arXiv230516803T}. As a consensus, 
collective effects in the dynamics are behind a robust route for designing efficient QBs, when compared with the parallel (non-interacting) charging of the system~\cite{gyhm2022quantum}. 
From the experimental point of view, the energy storing properties of a diversity of quantum systems have been exploited, with physical realizations in a number of architectures, 
such as nuclear magnetic resonance~\cite{PhysRevA.106.042601}, superconducting integrated circuits~\cite{Hu:21,batteries8050043}, metal complex~\cite{Cruz:22}, and an all-optical setup~\cite{PhysRevA.107.L030201}.

\begin{figure}[t!]
	\includegraphics[width=\linewidth]{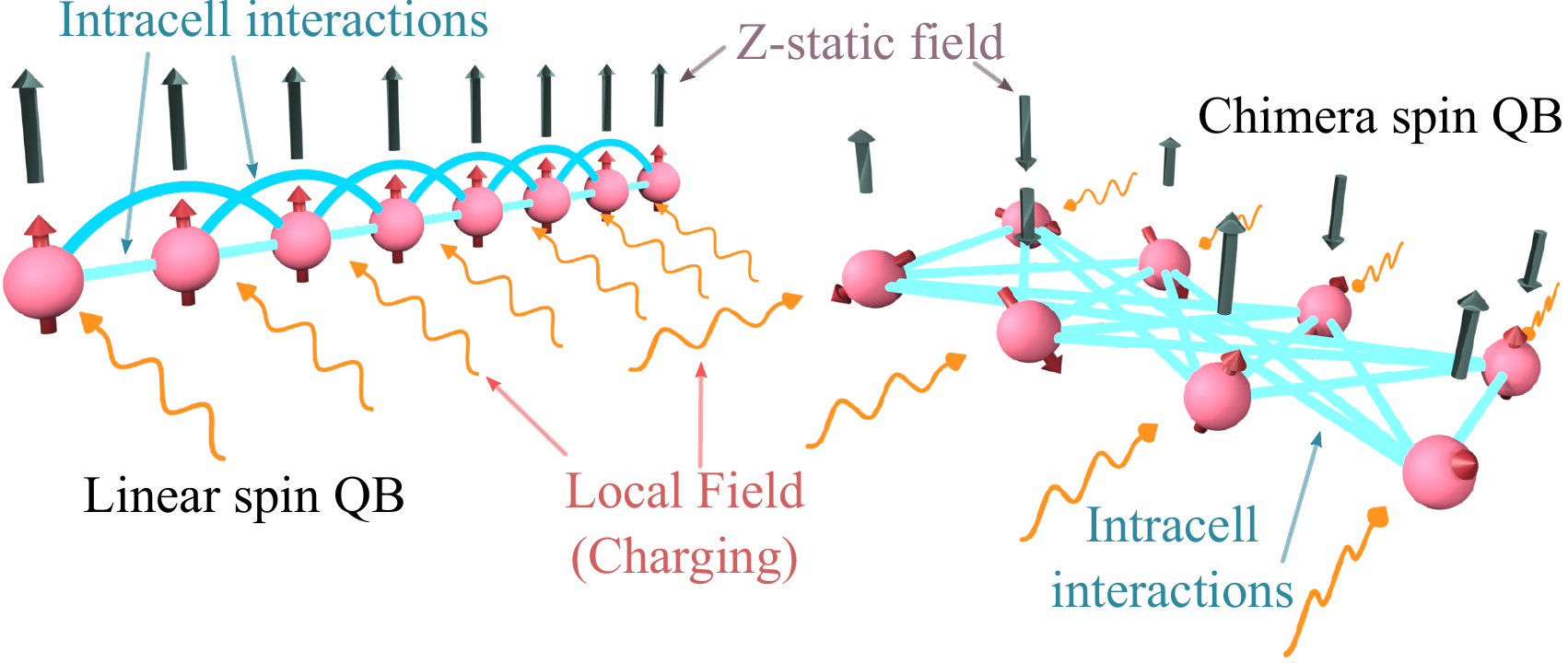}
	\caption{Two distinct schemes of QBs based on the transverse-field random Ising model are considered: (i) Ising couplings on a linear chain with up to next-to-nearest neighbor interactions (left) and 
		(ii) Ising couplings defined on a Chimera graph (right). For both models, the charging process is locally driven by an external field. The set of intracell interactions defines the kind of model adopted, 
		such that different phases can be obtained as a function of the disorder strength.}
	\label{schematic}
\end{figure}

Here, we aim at investigating quantum phases of matter that may be suitable for qubit-based QBs. Concretely, we will focus on the ergotropy and charging power of 
quantum devices subject to disordered couplings and external fields. As sketched in Fig.~\ref{schematic}, two distinct schemes for QBs based on the transverse-field random Ising model will be considered, 
with Ising couplings defined on a Chimera graph (for a physical realization, see e.g. Ref.~\cite{da2022localization}) and on a linear chain with up to next-to-nearest neighbor interactions~\cite{kjall2014many}. 
Analyses of disorder effects in the performance of QBs have been previously considered in the literature, such as in Refs.~\cite{ghosh2020enhancement,Andolina:19-2}. 
However, differently from previous works, where time-dependent interactions drive the charging process, we adopt permanent interactions, so that the QB charging is performed through external local fields only. 
Our approach is motivated by the fact that switching on/off interactions typically costs a significant amount of energy in comparison with the ergotropy provided by the QB~\cite{Cruz:22}. 
As a result, we obtain that the ergodic phase favors the QB performance, with disorder being detrimental to ergotropy due to the memory effects in the many-body localization (MBL) phase. 
Remarkably, when Anderson localization is available, we have a hybrid scenario between ergodic and MBL behaviors depending on the disorder strength. 
We also address the splitting of ergotropy into coherent and incoherent contributions, showing that a residual extractable work robust against dephasing can be stored as incoherent ergotropy~\cite{shi2022entanglement}. 

\section{Preliminaries}
\label{2}
We review now the concept of ergotropy and introduce the quantum spin systems adopted as models of QB in our work.

\subsection{Ergotropy}

The energy available to be extracted in the form of work in a QB depends on both the state $\rho$ of the quantum device and the reference Hamiltonian $H_{\mathrm{ref}}$ that sets the energy scale of the system. 
Given $\rho$ and $H_{\mathrm{ref}}$, the internal energy is provided by the expectation value $E(\rho,H_{\mathrm{ref}})=\tr(\rho H_{\mathrm{ref}})$. 
By keeping a constant entropy, work can be performed by/on the QB by unitary operations. In this scenario, we define {\it ergotropy} as the maximum energy extractable by realizing unitary operations on the system~\cite{Allahverdyan:04}. 
Therefore, ergotropy reads
\begin{equation}
\mathcal E(\rho,H_{\mathrm{ref}})=\max_U \left[ E(\rho,H_{\mathrm{ref}}) - E(U\rho U^\dagger,H_{\mathrm{ref}}) \right].
\end{equation}
The maximization over $U$ is achieved by driving the QB towards the {\it passive state} $\bar{\rho}\equiv U\rho U^\dagger$ associated with $\rho$, i.e., a state that cannot provide further energy by unitary operations. 
We can compute $\mathcal E(\rho,H_{\mathrm{ref}})$ by looking at the spectral decompositions of $\rho = \sum_{i=1}^{d} {p_{i} \ket{p_{i}}\bra{p_{i}}}$ and 
$H_{\mathrm{ref}} = \sum_{i=1}^{d} {\epsilon_{i} \ket{\epsilon_{i}}\bra{\epsilon_{i}}}$, where we assume the ordering $p_{1}\geq p_{2}\geq . . . \geq p_{d}$ and 
$\epsilon_{1}\leq \epsilon_{2}\leq . . . \leq \epsilon_{d}$, with $d$ being the dimension of the Hilbert space. 
Thus, ergotropy can be rewritten as~\cite{Allahverdyan:04} (see also Refs~\cite{Luiz:21,arjmandi2022enhancing})
\begin{align}
	\mathcal E(\rho,H_{\mathrm{ref}})=E(\rho,H_{\mathrm{ref}})-\sum_{i=1}^{d}{\epsilon_{i} p_{i}} . \label{eq1}
\end{align}
Notice that both the internal energy and the ergotropy depend on the reference Hamiltonian $H_{\mathrm{ref}}$. 
In the charging process of a QB, we also add a charging Hamiltonian $H_{\mathrm{ch}}(t)$, yielding a total driving Hamiltonian $H_{\mathrm{d}}(t) = H_{\mathrm{ref}} + H_{\mathrm{ch}}(t)$. 
Let us take $H_{\mathrm{ref}} = H_0 + H_{\mathrm{int}}$, where $H_0$ is a local Hamiltonian and $H_{\mathrm{int}}$ contains the native system interactions. 
Then, any additional cost to charge the battery comes from the charging Hamiltonian $H_{\mathrm{ch}}(t)$. More precisely, the cost to turn-on and turn-off the term $H_{\mathrm{ch}}(t)$ 
should be considered in the energy balance of the device, which may lead to non-efficient quantum batteries~\cite{Cruz:22}. Here, to avoid the cost of engineering non-native interactions, 
we focus on a charging process driven only by external local fields.

\subsection{QB Models} \label{Sec:Model}

As sketched in Fig.~\ref{schematic}, we consider two distinct models of QB, with many-body localization regimes properly characterized. First, let us consider an $N$-particle random spin-1/2 Ising linear chain with on-site magnetic static field $h_k$. 
The native interactions are provided by nearest neighbor ($J^{(1)}_{k}$) and next-to-nearest neighbor ($J^{(2)}_{k}$) couplings. The reference Hamiltonian of the system reads
\begin{align}
H_{\mathrm{ref}}^{\mathrm{Isi}}= \sum_{k=1}^{N} {h_{k}\sigma_{k}^{z}}-\sum_{k=1}^{N-1} {J^{(1)}_{k} \sigma_{k}^{x} \sigma_{k+1}^{x}}+ \sum_{k=1}^{N-2} J^{(2)}_{k}{\sigma_{k}^{x} \sigma_{k+2}^{x}},  \label{eq2}
\end{align}
where $\sigma^{\alpha}$ $(\alpha\!=\!x,y,z)$ are the standard Pauli matrices.
The QB is charged according to a transverse local field through a charging Hamiltonian
\begin{align}
H_{\mathrm{ch}}= \sum_{k}^{N}{\Omega_{k}\sigma_{k}^{x}}. \label{Eq:CharginHamiltonian}
\end{align}
This driving field can be implemented, for instance, through an oscillating small radio-frequency magnetic field of intensity $|B_{\mathrm{rf}}| \propto \Omega_{k} $, as in a nuclear magnetic resonance setup~\cite{Sarthour:Book}. 
We assume local fields uniformly distributed over the chain, with $h_{k}=h$ and $\Omega_{k}=\Omega$ constant couplings. In this model, the random nature of the system relies on the interactions between nearest neighbors given 
by $J^{(1)}_{k}$, as selected from a uniform distribution in the interval $[J_{0} (1 - \delta) , J_{0} (1+ \delta)]$, where $J_{0}$ sets the energy scale of the system and  $\delta$ is a dimensionless parameter that determines the degree of disorder. 
Concerning the next-to-nearest neighbor coupling $J^{(2)}_{k}$, it will also be taken as a constant chosen as a function of $J_{0}$. For the charging process we assume $J_{0} = \Omega$.

The localization properties of random spin-1/2 Ising chains have been widely studied~\cite{kjall2014many}. In particular, as the next-to-nearest neighbor interaction is turned off ($J^{(2)}_{k}=0$), the model reduces to the well known 
nearest neighbor transverse field Ising chain, such that an arbitrary amount of disorder ($\delta > 0$) brings the system into an Anderson localization regime~\cite{vznidarivc2008many}. However, when $J^{(2)}_{k} \neq 0$, the degree of disorder needs to 
be strong enough to drive the system to many-body localization (MBL), otherwise the system remains in an ergodic phase~\cite{vznidarivc2008many,pal2010many,luitz2015many}. 
Unless specified otherwise, we refer to the various phases of this model by adopting the specific values for the couplings: (i) Anderson localized: $J^{(2)}_{k}=0$ and $\delta=1$; (ii) Ergodic phase: $J^{(2)}_{k} = h/2 = 0.3 J_{0}$ and $\delta=1$, 
and (iii) MBL phase: $J^{(2)}_{k}=h/2=0.3 J_{0}$ and $\delta=5$.

It is worth mentioning that this model was previously considered to investigate localization effects on QB~\cite{Andolina:19-2}. However, in Ref.~\cite{Andolina:19-2}, the authors assume the QB is charged through {\it turned-on} 
interactions between the quantum cells of the battery during charging time. Then, all figures of merit are evaluated with respect to the free part of the Hamiltonian, i.e., the on-site applied field. As previously mentioned, 
the energy cost of switching on/off the interactions might be detrimental to the performance of the QB, with the energy cost to charge the battery much larger than the ergotropy itself~\cite{Cruz:22,Hovhannisyan:20}. 
Here, we analyze the energy cost of the charging process in Appendix~\ref{Apendix:EnergyCost}, showing that the charging through a local field only is an energy friendly choice.

As our second QB model, we consider random Ising interactions and local fields defined on the 8-spin Chimera unit cell, as sketched in Fig.~\ref{schematic}. 
The reference Hamiltonian reads
\begin{align}
H_{\mathrm{ref}}^{\mathrm{Chi}}=\sum_{k} {h_{k} \sigma_{k}^{z}}+\sum_{\langle k,j\rangle \in \Ccal_{\mathrm{Chi}} } {J_{kj} \sigma_{k}^{z} \sigma_{j}^{z}},\label{eq4}
\end{align}
where the $\langle k,j\rangle \in \Ccal_{\mathrm{Chi}}$ refers to a sum over the connections of the Chimera graph. Concerning the charging Hamiltonian $H_{\mathrm{ch}}$, it will be defined through transverse local fields, as  
in Eq.~\eqref{Eq:CharginHamiltonian}, with constant couplings $\Omega_{k}=\Omega$. 
For this system, we consider random values for both $h_{k}$ and $J_{kj}$ from a uniform distribution $[- \delta J_0 , \delta J_0]$, with $\delta$ being the disorder amplitude, such that $\delta J_0$ is the maximum admissible value for the coupling strength.
Taking into account the characterization of the MBL phase transition as in Ref.~\cite{da2022localization}, with $J_{0}=\Omega$, we will adopt the specific values for the disorder strength: (i) Ergodic phase: $\delta= 2$, such that $\delta J_{0}=2\Omega$ and (ii) MBL phase: $\delta= 6$, such that $\delta J_{0}=6\Omega$.
It is worth mentioning that, while the Anderson localized phase is well defined for the random Ising chain in the absence of next-to-nearest-neighbor interactions (i.e., when $J_{2}=0$)~\footnote{since by using a Jordan-Wigner transformation~\cite{sachdev1999quantum} the model can be mapped into a non-interacting fermionic model, for which any arbitrary degree of disorder is able to localize the system~\cite{kjall2014many}}, 
we do not expect that the chimera topology will exhibit Anderson localization for the regime of couplings considered here.

By disregarding decoherence effects, the dynamics of the QB is governed by the unitary evolution operator $U(t)$ as $\rho(t)= U(t) \rho(0) U^{\dagger}(t)$. We assume the device starts from its fully empty state $\rho(0) = \ket{G}\bra{G}$, where $\ket{G}$ is the ground state of the reference Hamiltonian for each model (empty energy state of the battery). Given the randomness of local fields and interaction strengths, we will have a distinct reference Hamiltonian $H_{n,\mathrm{ref}}$ for each disorder realization. 
Therefore, it is expected that the ergotropy will change for different choices of the random parameters. For this reason, to define a fair figure of merit to evaluate the QB performance, we will consider the average ergotropy over $N_{\mathrm{r}}$ realizations of disorder~\cite{arjmandi2022enhancing}
\begin{align}
\bar{\mathcal E}(t)=\frac{1}{N_{\mathrm{r}}} \sum_{n=1}^{N_{\mathrm{r}}}\frac{\mathcal E(\rho_{n}(t),H_{n,\mathrm{ref}})}{\mathcal{E}_{n}^{\mathrm{max}}} , \label{eq6}
\end{align}
where $\mathcal E(\rho_{n}(t),H_{n,\mathrm{ref}})$ is the time-dependent ergotropy associated with the evolved state $\rho_{n}(t)$ for the $n$-th realization. 
We also define $\mathcal{E}_{n}^{\mathrm{max}}= \tr(\rho_{n,\mathrm{fc}} H_{n,\mathrm{ref}}) - \tr(\rho_{n,G} H_{n,\mathrm{ref}})$ as the maximum capability of the battery, i.e., the ergotropy stored in its fully charged state $\rho_{n,\mathrm{fc}}$, 
with $\rho_{n,G}$ denoting the density operator for the ground state of $H_{n,\mathrm{ref}}$. 
Because we assume pure states (unitary evolution), $\rho_{n,\mathrm{fc}}$ is the eigenstate of $H_{n,\mathrm{ref}}$ with the highest eigenvalue~\cite{Allahverdyan:04,Santos:20c}. In this way, we can measure the charge in the QB 
with the reference Hamiltonian $H_{n,\mathrm{ref}}$ for a given realization $n$, with respect to the maximally charged state $\rho_{n,\mathrm{fc}}$. Throughout this work we show the results for the case with $N_{r}=100$, but it is worth saying that we do not observe any significant change by taking $N_{r}>100$.

\section{Results and discussion} \label{3}

\subsection{Ergotropy and charging power}

\begin{figure}[t!]
	\includegraphics[width=\linewidth]{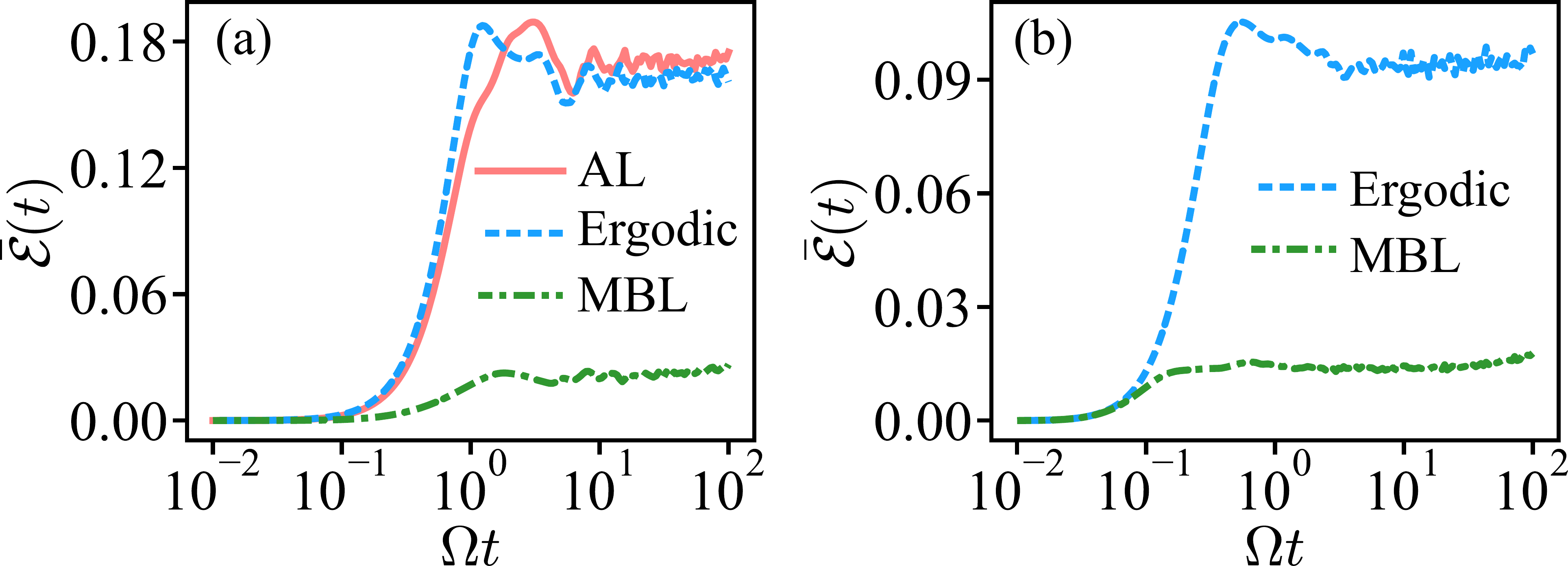}
	\caption{The dynamics of average ergotropy for (a) the $N=8$ random Ising chain with next-to-nearest neighbors and (b) the random Ising model on the 8-spin Chimera graph. The number of realizations is taken as $N_{\mathrm{r}}=100$.  
	The Hamiltonian parameters for each phase are set as described in Sec.~\ref{Sec:Model}.}
	\label{erg}
\end{figure}

Now, we present our main results. First of all, we evaluate the ergotropy by charging the battery in the different phases aforementioned. 
In Fig.~\ref{erg}, we show the dynamics of the average ergotropy for both models considered in our work, where we conclude that the MBL phase turns out to be a poor choice for a QB in the sense of energy storage. 
This can be explained by the emergence of memory effects attributed to the MBL phase with respect to the initial state~\cite{nandkishore2015many,da2022localization}, which in turn prevents the QB from a significant evolution 
with respect to its ground state. If the dynamics starts from a different state, an enhancement of the charging performance is expected, but it requires the initialization step by using a different reference Hamiltonian. 
For example, in Ref.~\cite{Andolina:19-2}, the authors considered the initial state as the lowest energy level of the non-interacting quantum cells, which will demand an extra cost as we turn on the interactions throughout the charging process.

\begin{figure}[t!]
	\includegraphics[width=\linewidth]{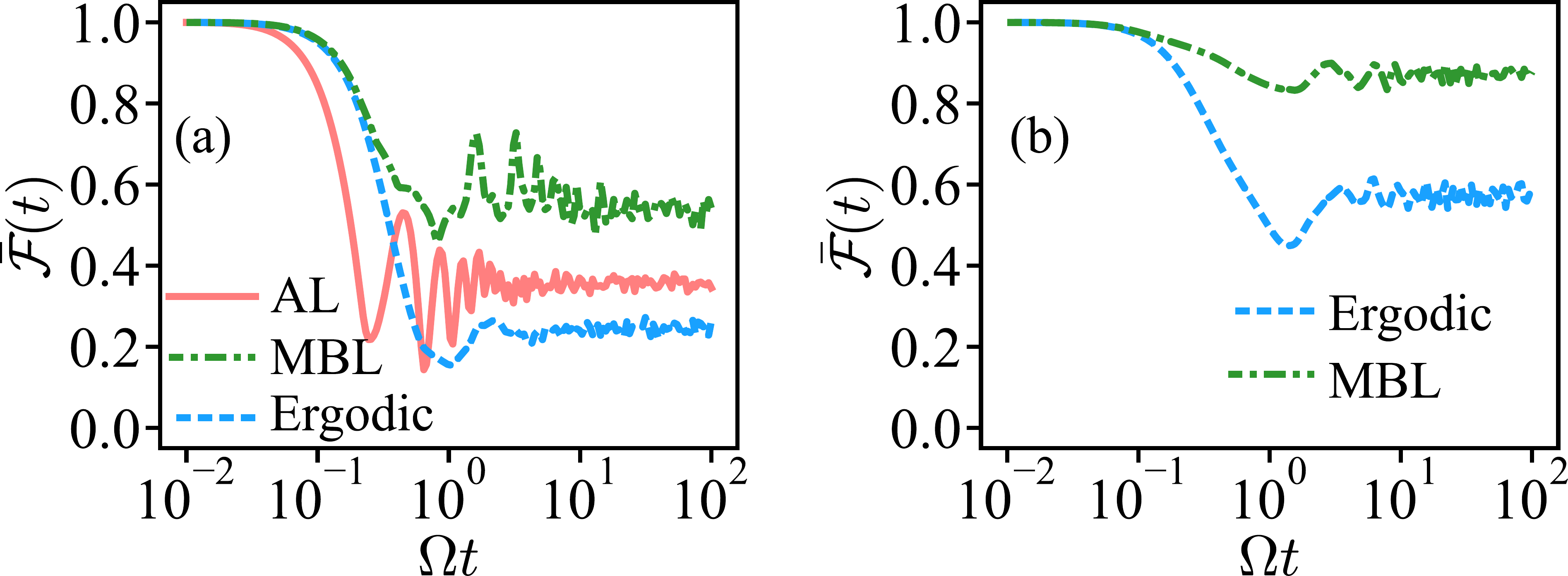}
	\caption{Average fidelity $\bar{\mathcal F}(t)$ as defined by Eq.~(\ref{fidelity0}) for (a) the $N=8$ random Ising chain with next-to-nearest neighbors and (b) the random Ising model on the 8-spin Chimera graph. 
	The number of realizations is taken as $N_{\mathrm{r}}=100$.  
	The Hamiltonian parameters for each phase are set as described in Sec.~\ref{Sec:Model}.}
	\label{fidelity}
\end{figure}

\begin{figure}[t!]
	\includegraphics[width=\linewidth]{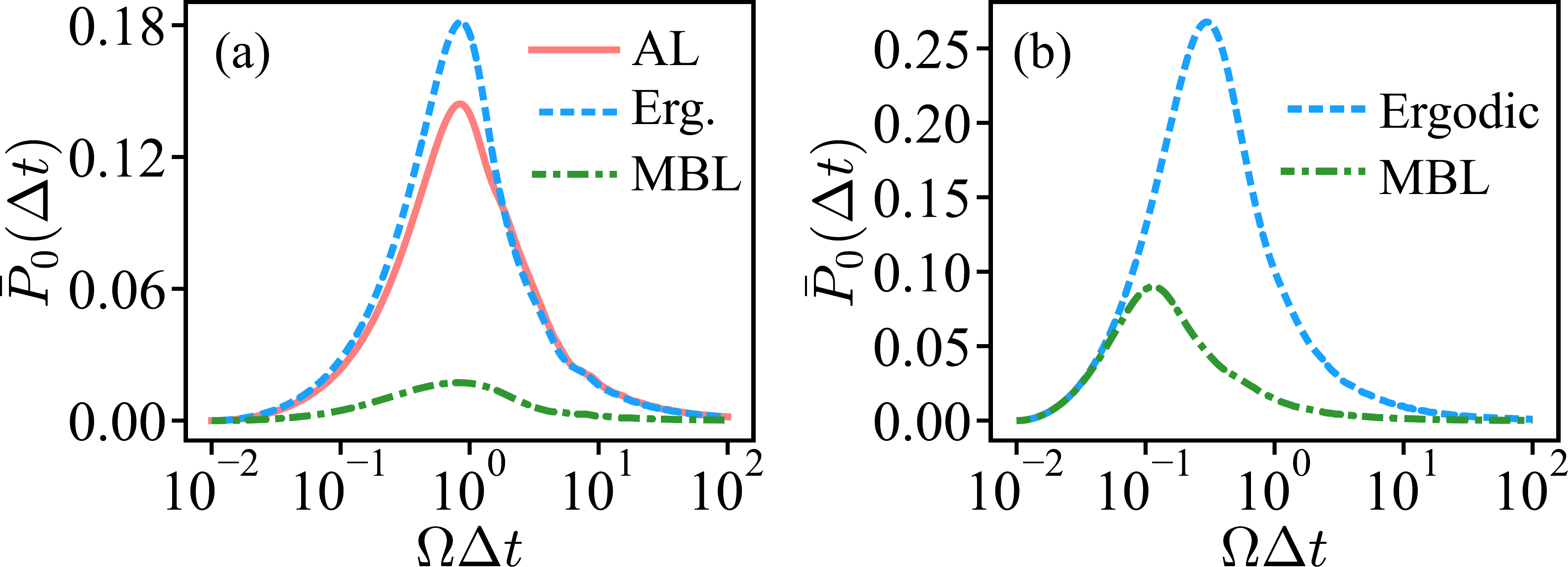}
	\caption{The dynamics of averaged charging power for (a) the $N=8$ random Ising chain with next-to-nearest neighbors and (b) the random Ising model on the 8-spin Chimera graph. 
	The number of realizations is taken as $N_{\mathrm{r}}=100$.  
	The Hamiltonian parameters for each phase are set as described in Sec.~\ref{Sec:Model}.}
	\label{pow}
\end{figure}

Differently from the MBL phase, where the ``frozen" dynamics handicaps the energy charging, the battery presents a charging enhancement in either the ergodic or Anderson phases. In the case of the ergodic phase, 
such a result can be explained by its delocalization behavior, which makes the ergodic phase more suitable for ergotropy deposition. Notice that the Anderson phase follows a similar pattern as obtained in the ergodic phase for the chosen disorder strength, 
also showing a 
remarkable difference with respect to the MBL phase in terms of ergotropy. 
Due to the absence of strong memory effects typical of the MBL phase, the dynamics of QB in the Anderson phase is not frozen so that it shows a suitable performance. 
Moreover, for a wide time interval, which is around the range $[10^0,10^2]$, the Anderson phase presents a performance even better than the ergodic case. 
We can reach a more precise view at the above discussion by analyzing the memory effects of MBL phase and its poor performance for ergotropy storage. To this end, we need to know how the evolved state of the QB looks like 
when compared to its initial ground state. More resemblance to the initial state can be inferred as a sign of retaining information about the past and, consequently, indicating a memory effect. Therefore, we consider the average fidelity, 
taken over $N_{\mathrm{r}}$ disorder realizations, between the time-dependent density operator of the QB and its initial ground state, which reads
\begin{align}
\bar{\mathcal F}(t)=\frac{1}{N_{\mathrm{r}}} \sum_{n=1}^{N_{\mathrm{r}}} \mathcal F (\rho_{n}(t),\rho_{n,G}), \label{fidelity0}
\end{align}
where the fidelity of two general density matrices $\rho$ and $\sigma$ is defined by $\mathcal F (\rho,\sigma)=\mathrm{Tr} \sqrt{\sqrt{\rho}\sigma \sqrt{\rho}}$.
As shown in Fig.~\ref{fidelity} the highest fidelity over the considered charging time belongs to the MBL phase, implying the minimum change in the state of the QB with respect to its initial preparation. This fact is more visible for the Chimera model.
Let us now analyze the time-averaged charging power of the QB as a function of time. Notice that, as shown in from Fig.~\ref{erg}, the maximum instantaneous amount of ergotropy is achieved for different instants of time for each phase considered. 
Then, it is meaningful to consider the time-averaged charging power in the parametrized time interval $\Omega \Delta t = \Omega (t - t_0)$ (assuming here $t_0=0$). In this context, by using the notation introduced in Eq.~(\ref{eq6}), we can introduce the dimensionless power parameter
\begin{align}
\bar{P}_{0}(\Delta t)=\frac{\bar{\mathcal E}(t) - \bar{\mathcal E}(0)}{\Omega \Delta t}. \label{power0}
\end{align}
Notice that $\bar{P}_{0}(\Delta t)$ is directly related to the average power $P(\Delta t)$. Indeed, $P(\Delta t)$ can be written in terms of the average of the maximum ergotropy capability of the battery $\bar{\mathcal{E}}^{\mathrm{max}}=\sum_{n=1}^{N_{\mathrm{r}}}\mathcal{E}_{n}^{\mathrm{max}}/N_{\mathrm{r}}$, the charging field amplitude $\Omega$, and the parameter $\bar{P}_{0}(\Delta t)$ as $P(\Delta t) = \Omega\bar{\mathcal{E}}^{\mathrm{max}} \bar{P}_{0}(\Delta t) $. Through the above definition, we can collect information about the charging performance (including the optimal charging time window $\Delta t_{\mathrm{opt}}$) defined over the entire time-domain of the charging process, which makes this analysis more robust than simply finding the optimal charging time for the battery.
For both models, as it can be seen from Fig.~\ref{pow}, the QB in the ergodic phase is more powerful than the other phases for 
the range of time considered. Then, while the Anderson phase may allow for storing more energy than the ergodic phase for a specific time range, its charging power is not as efficient as the ergodic QB.

For completeness of our discussions, it is worth mentioning that different behaviors for the ergotropy and phase-dependent battery performance can be drastically modified for other kinds of fields chosen to charge the battery. We identified that such a change is mainly due to the charging-field induced change in the localization transition of the system, then leading to a significant effect on the nature of the models and their quantum phases (for more details, see Appendix~\ref{Ap:CharEff}.) In addition, in Appendix~\ref{Ap:NNN} we included a brief analysis of the expected negative effects of random next-to-nearest neighbor interaction ($J_{2}$). As a conclusion of this case, we observe that a small reduction in the quality of charging process is expected when introducing this extra randomness, such that additional efforts to engineer random next-to-nearest neighbor couplings $J_{2}$ do not help the charging process.

\begin{figure}[t!]
	\includegraphics[width=\linewidth]{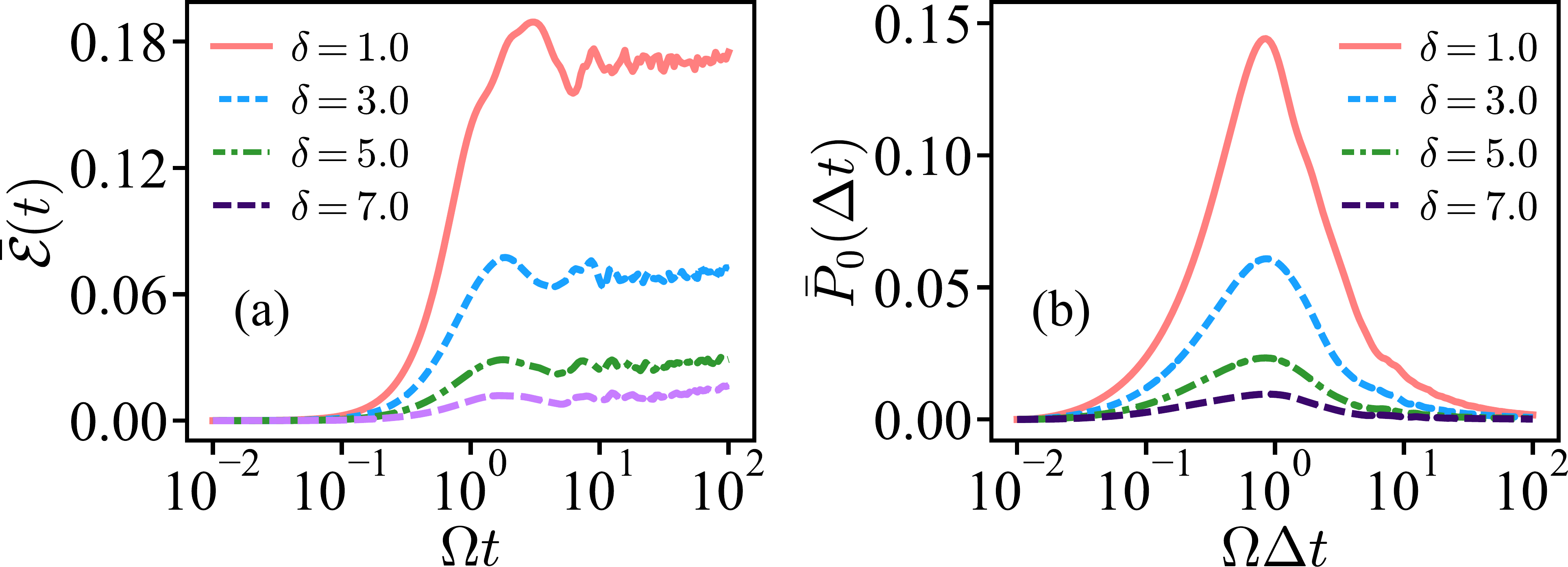}
	\caption{The dynamics of (a) average ergotropy and (b) power for the $N=8$ random Ising chain in the Anderson phase for different degrees of disorders. 
	The number of realizations is taken as $N_{\mathrm{r}}=100$.  
	The Hamiltonian parameters for each phase are set as described in Sec.~\ref{Sec:Model}.}
	\label{DisprPower}
\end{figure}

So far, our results suggest that the disorder strength seems to be the main determinant for ergotropy and power rather than the quantum (ergodic or Anderson) phase the system actually is. 
In fact, the charging of the QB in the Anderson localized and ergodic phases are performed by using an equal degree of disorder ($\delta=1$), 
while the MBL phase requires stronger disorder. The role played by disorder can be confirmed by considering the Anderson phase for different degrees of randomness. 
This is shown in Fig.~\ref{DisprPower}, where we have plotted ergotropy and dimensionless power in the Anderson phase for the Ising linear chain for different values of $\delta$. Notice that the Anderson phase then 
interpolates between the ergodic and MBL behaviors as we increase the strength of disorder. 

\subsection{Incoherent and coherent ergotropy}

Let us now split ergotropy into its \textit{coherent} and \textit{incoherent} contributions~\cite{shi2022entanglement}. 
The generation of quantum superpositions (coherence) in the energy basis during the charging leads to a non-zero amount of coherent ergotropy. On ther other hand, part of the total ergotropy is stored as incoherent ergotropy. 
In particular, the incoherent part of the ergotropy is relevant in our discussion because it takes into account the ``residual" ergotropy of the \textit{dephased} state $\rho^{D}$ in energy eigenbasis of the reference Hamiltonian, 
with $\rho^{D}=\sum_{i}^{d}{\bra{\epsilon_{i}}\rho\ket{\epsilon_{i}}\ket{\epsilon_{i}}\bra{\epsilon_{i}}}$. Physically speaking, it quantifies the total amount of ergotropy that remains in the system, even in absence of an external 
charging field under a dephasing process. Therefore, we can identify this amount of remaining ergotropy as related to the robustness of a QB against its \textit{self-discharging process}~\cite{Santos:21b}. 

\begin{figure}[t!]
	\includegraphics[width=\linewidth]{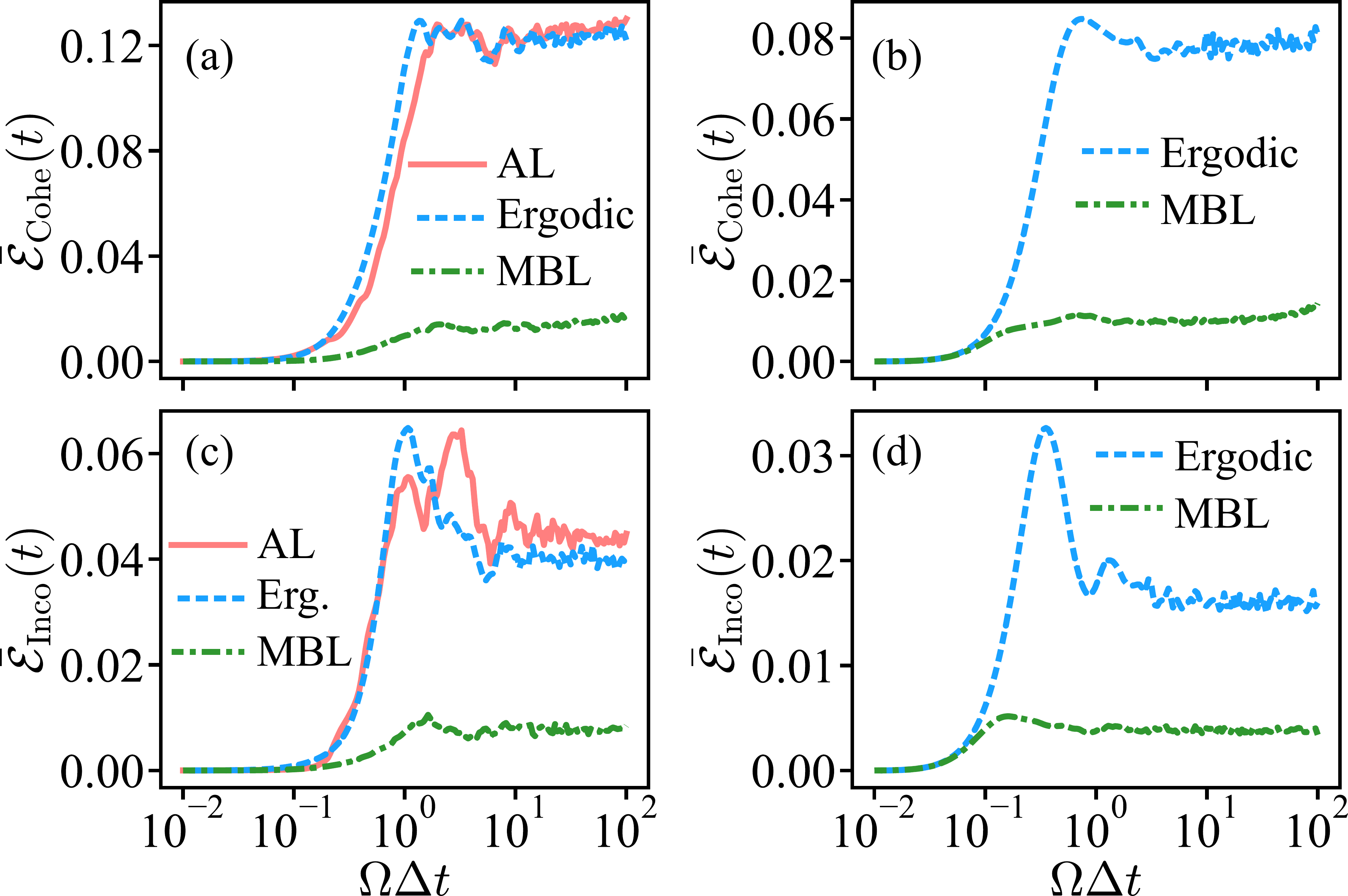}
	\caption{The dynamics of coherent and incoherent ergotropy for (a), (c) the $N=8$ random Ising chain with next-to-nearest neighbors and (b), (d) the random Ising model on the 8-spin Chimera graph. 
	The number of realizations is taken as $N_{\mathrm{r}}=100$.  
	The Hamiltonian parameters for each phase are set as described in Sec.~\ref{Sec:Model}.}
	\label{work}
\end{figure}

Inspired by Ref.~\cite{shi2022entanglement}, we introduce the average normalized incoherent and coherent ergotropy, respectively, as
\begin{align}
	\bar{\mathcal E}_{\mathrm{Inco}}(t)&=\frac{1}{N_{\mathrm{r}}} \sum_{n=1}^{N_{\mathrm{r}}}\frac{\mathcal E(\rho_{n}^{D}(t),H_{n,\mathrm{ref}})}{\mathcal{E}_{n}^{\mathrm{max}}} , \label{eq7}\\
	\bar{\mathcal E}_{\mathrm{Cohe}}(t)&=\frac{1}{N_{\mathrm{r}}} \sum_{n=1}^{N_{\mathrm{r}}}\frac{\mathcal E(\rho_{n}(t),H_{n,\mathrm{ref}})- \mathcal E(\rho_{n}^{D}(t),H_{n,\mathrm{ref}})}{\mathcal{E}_{n}^{\mathrm{max}}} . \label{eq8}
\end{align}
Fig.~\ref{work} shows the time evolution of $\bar{\mathcal E}_{\mathrm{Inco}}(t)$ and $\bar{\mathcal E}_{\mathrm{Cohe}}(t)$  for different phases of the models considered in our work. 
Given the good qualitative agreement between the behavior of the total extractable energy and the its coherent and incoherent parts, we can safely state that the performance of the QB 
is mainly related to the coherence generated during its charging, since the extractable work is mainly stored as coherent ergotropy. However, by comparing Fig.~\ref{erg} and Fig.~\ref{work}, 
it can be inferred that, even though the total ergotropy is mostly consisted of coherent part, there remains a nonzero amount of incoherent ergotropy during charging process. 
This can be useful in a real setup since, as discussed above, the incoherent ergotropy is robust against dephasing channels, since its corresponding state is already dephased in the basis of eigenenergies. It is worth mentioning that the incoherent ergotropy can be wasted under dissipation processes through population relaxation. Nonetheless, one may find a situation in which pure dephasing is the main dominant process, namely for quantum dots~\cite{krummheuer2002theory,fan1998pure,besombes2001acoustic,vagov2004nonmonotonous} or trapped ions~\cite{turchette2000decoherence,myatt2000decoherence}, resulting in potential usefulness of the incoherent ergotropy.

\begin{figure}[t!]
	\includegraphics[width=\linewidth]{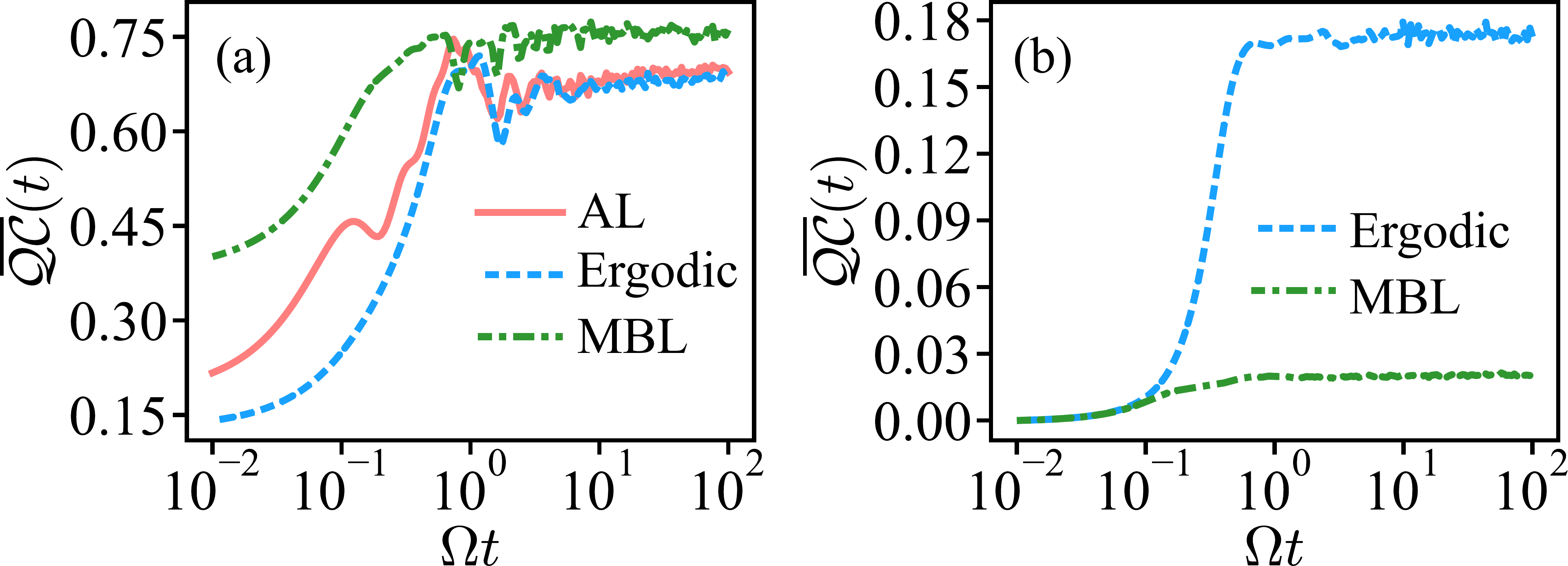}
	\caption{Average quantum coherence for (a) the $N=8$ random Ising chain with next-to-nearest neighbors and (b) the random Ising model on the 8-spin Chimera graph. The number of realizations is taken as $N_{\mathrm{r}}=100$.  
	The Hamiltonian parameters for each phase are set as described in Sec.~\ref{Sec:Model}.}
	\label{coh}
\end{figure}

We can now analyze the average normalized coherence ${\overline{\mathcal Q \mathcal C}}(t)$ generated for each dynamics as quantified by the \textit{$l_{1}$}-norm, which is defined as
\begin{align}
{\overline{\mathcal Q \mathcal C}}(t)=\frac{1}{N_{\mathrm{r}}} \sum\nolimits_{n=1}^{N_{\mathrm{r}}} \left[\frac{1}{C_{max}} \sum\nolimits _{i,j\neq i} |\rho_{ij}|\right], \label{coherence0}
\end{align}
In Eq.~(\ref{coherence0}), the term in ``brackets" is the \textit{$l_{1}$}-norm coherence of the density matrix for each realization $n$, where $C_{max}$ denotes the coherence of the quantum state $(\ket{e}+\ket{g})/\sqrt{2}$, 
which is the maximally coherent state in the local basis $\{\ket{e},\ket{g}\}$. The results for ${\overline{\mathcal Q \mathcal C}}(t)$ are shown in Fig.~\ref{coh} for both the Ising chain and the Chimera graph.  
Comparing Figs.~\ref{erg}{\color{red}b} and~\ref{coh}{\color{red}b}, for the Chimera graph, quantum coherence can be regarded as a suitable figure of merit of the stored ergotropy, at least qualitatively. 
As time passes by, both quantities sharply increase when the QB is in the ergodic phase but they are small for the MBL phase. This is not the case for the Ising chain. 
The model contains some amount of initial quantum coherence but localized phases reach more coherence in comparison with the ergodic phase. 
The relation between generation of ergotropy and coherence is then model-dependent ,i.e., different patterns of interactions may imply in different qualitative relations.

\section{Conclusions}
\label{4}
We have investigated quantum phases of disordered spin systems as a model of qubit-based QBs. Differently from previous investigations, we have focused on the local ergotropy injection through external fields only. This is 
due to the low energy demanding property of this scheme in comparison with the time-dependent switchable interacting systems. 
For the models considered in our work, ergodicity has shown to improve the QB performance, with disorder being detrimental to ergotropy due to the memory effects of the discharged state in the MBL phase. 
In the case of the linear Ising chain, where Anderson localization is induced for any amount of disorder as the next-to-nearest interaction is turned off, we have shown the presence of a hybrid scenario interpolating 
between ergodic and MBL behaviors depending on the disorder strength. 
It is worth highlighting here that, while the MBL phase is attractive for time-dependent interacting QBs~\cite{Andolina:19-2}, it develops a poor performance for the local charging scheme. 
Concerning quantum coherence, we identified that most of ergotropy is stored as coherent ergotropy, which provides the quantum character of the energy storing device. 
On the other hand, we also obtained a nonvanishing amount of incoherent ergotropy, which indicates robustness of the storage capability against the self-discharging of QBs due to dephasing 
(see, e.g., Refs.~\cite{Santos:21b} and~\cite{shi2022entanglement}). Further investigations of lattice topologies and quantum phases in disordered systems as well as experimental implementations are left for a future analysis.  

\begin{acknowledgements}
M.S.S. is supported by Conselho Nacional de Desenvolvimento Cient\'{\i}fico e Tecnol\'ogico (CNPq) (307854/2020-5).
A.C.S. acknowledges the financial support of the São Paulo Research Foundation (FAPESP) (Grant No. 2019/22685-1). This research is also supported in part by Coordena\c{c}\~ao de Aperfei\c{c}oamento de Pessoal de N\'{\i}vel Superior (CAPES) (Finance Code 001) and by the Brazilian National Institute for Science and Technology of Quantum Information (INCT-IQ). 
\end{acknowledgements}

\appendix

\section{Energy cost of switching on/off interactions} \label{Apendix:EnergyCost}
\label{appA}
In this section, we analyze the energy cost of turning on/off the intracell interactions for both models of QB considered in our work. 
For the spin-$1/2$ Ising chain with nearest and next-to-nearest neighbor interactions we have $H_{\mathrm{ref}}^{\mathrm{Isi}}=H_{0}^{\mathrm{Isi}}+H_{\mathrm{int}}^{\mathrm{Isi}}$, where
\begin{align}
H_{0}^{\mathrm{Isi}}&=h \sum_{k=1}^{N} {\sigma_{k}^{z}} , ~~
H_{\mathrm{int}}^{\mathrm{Isi}}=-\sum_{k=1}^{N-1} {J_{k} \sigma_{k}^{x} \sigma_{k+1}^{x}}+J_{2} \sum_{k=1}^{N-2} {\sigma_{k}^{x} \sigma_{k+2}^{x}}.\label{HIl}
\end{align}
For the Ising couplings defined on the Chimera graph, we also have $H_{\mathrm{ref}}^{\mathrm{Chi}}=H_{0}^{\mathrm{Chi}}+H_{\mathrm{int}}^{\mathrm{Chi}}$ where
\begin{align}
H_{0}^{\mathrm{Chi}}=\sum_{k} {h_{k} \sigma_{k}^{z}}, ~~
H_{\mathrm{int}}^{\mathrm{Chi}}=\sum_{kj} {J_{kj} \sigma_{k}^{z} \sigma_{j}^{z}}.\label{HIC}
\end{align}
We introduce the energy cost $\mathcal C_{\mathrm{int}}$ to implement the interactions as the difference between the norm of the reference Hamiltonian including interactions ($H_{\mathrm{ref}}$) 
and the norm of the reference Hamiltonian without interactions ($H_{0}$). Mathematically, we have
\begin{align}
\mathcal C_{\mathrm{int}}=\mathcal N (H_{\mathrm{ref}}) - \mathcal N (H_{0}), \label{hsnormint}
\end{align}
where $\mathcal N (A)=\sqrt{Tr(A^{\dagger} A)}$ is the Hilbert-Schmidt norm of a given (finite-dimensional) matrix $A$. Our definition based on the Hilbert-Schmidt norm is motivated by 
previous works showing that such a quantity is a good measure of energy cost for quantum control in different contexts~\cite{Santos:15,Coulamy:16,Campbell-Deffner:17,Santos:18-b}, 
including experiments~\cite{Hu:18,Santos:20b}. More precisely, it is related to a measure of the thermodynamic cost of an arbitrary unitary quantum evolution~\cite{Deffner:21}. 
In this way, one can show that, for the Ising chain, the cost reads 
\begin{align}
\mathcal C^{\mathrm{Isi}}_{\mathrm{int}}=16 \left(-2 \sqrt{4 h^{2}} + \sqrt{8 h^{2} + 6 J_{2}^{2} + \sum_{k}{J_{k}^{2}}}\right), \label{hsnormlinear}
\end{align}
while, for the Chimera graph, we have
\begin{align}
\mathcal C^{\mathrm{Chi}}_{\mathrm{int}}=16 \left(-\sqrt{\sum_{k} {h_{k}^{2}}} + \sqrt{\sum_{k} {h_{k}^{2} + \sum_{kj} {J_{kj}}^{2}}}\right). \label{hsnormlinearChi}
\end{align}
In our work, interactions are always present and we only turn-on and turn-off the charging field. Then, we define the energy cost $\mathcal C_{\mathrm{ch}}$ of charging with local chargers as the difference between the norm of the 
driving Hamiltonian (including reference Hamiltonian and any local charging fields) and the norm of the reference Hamiltonian. Mathematically, we write
\begin{align}
\mathcal C_{\mathrm{ch}}=\mathcal N (H) - \mathcal N (H_{\mathrm{ref}}), \label{hsnormcha}
\end{align}
where $H=H_{\mathrm{ref}}+H_{\mathrm{ch}}$ is the total driving Hamiltonian. Again, one can simply obtain the explicit form of this cost for the two models
\begin{align}
\mathcal C^{\mathrm{Isi}}_{\mathrm{ch}}&=16 \left( \sqrt{\Xi^2 + 8 \Omega^{2}}- |\Xi|\right), \label{normchargising} \\
\mathcal C^{\mathrm{Chi}}_{\mathrm{ch}}&=16 \left(\sqrt{\Lambda^2 + 8 \Omega^{2}} -|\Lambda|\right). \label{normchargchimera}
\end{align}
with $\Xi^2 = 8 h^{2} + 6 J_{2}^{2} + \sum_{k}{J_{k}^{2}}$, and $\Lambda^2 = \sum_{k} {h_{k}^{2}} + \sum_{kj} {J_{kj}}^{2}$.

\begin{figure}[t!]
\includegraphics[width=\linewidth]{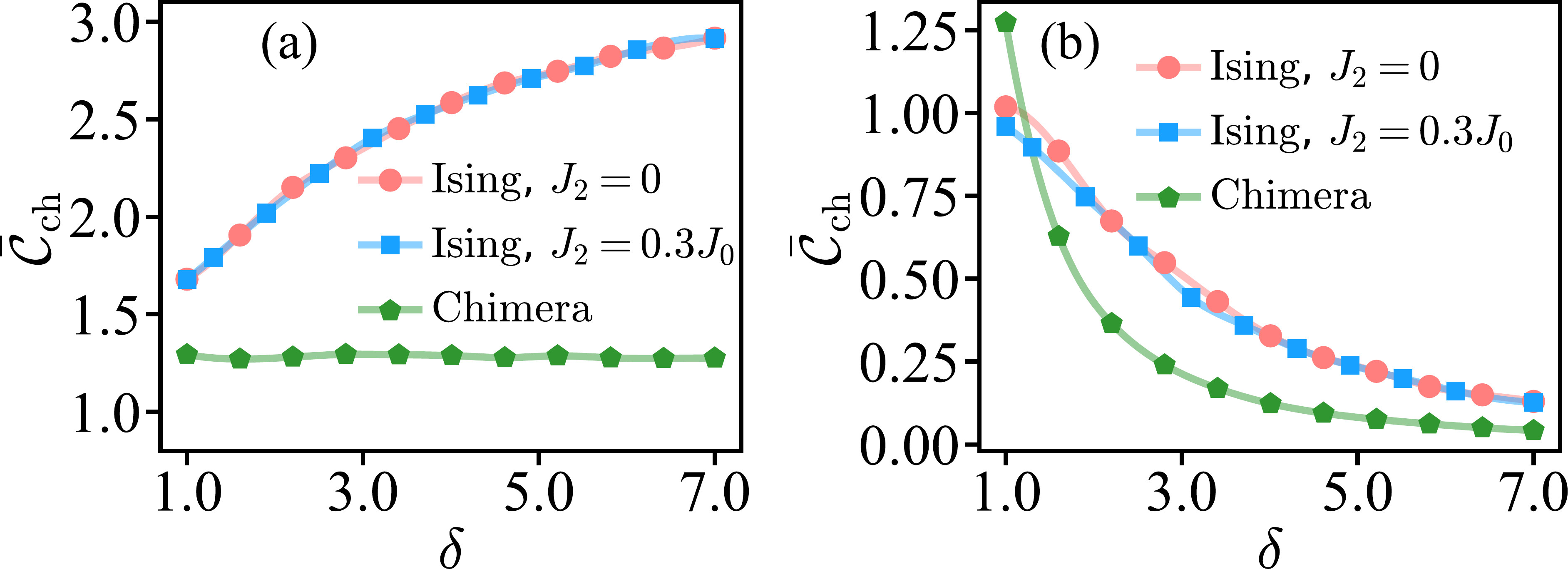}
\caption{Cost of turning on/off (a) intracell interactions, and (b) local charging fields as function of the degree of disorder for the $N=8$ random Ising chain with next-to-nearest neighbors and 
the random Ising model on the 8-spin Chimera graph. The number of realizations is taken as $N_{\mathrm{r}}=100$.  
	The Hamiltonian parameters for each phase are set as described in Sec.~\ref{Sec:Model}.}
\label{cost-dis}
\end{figure}

\begin{figure*}[t!]
	\centering
	\includegraphics[width=1.7\columnwidth]{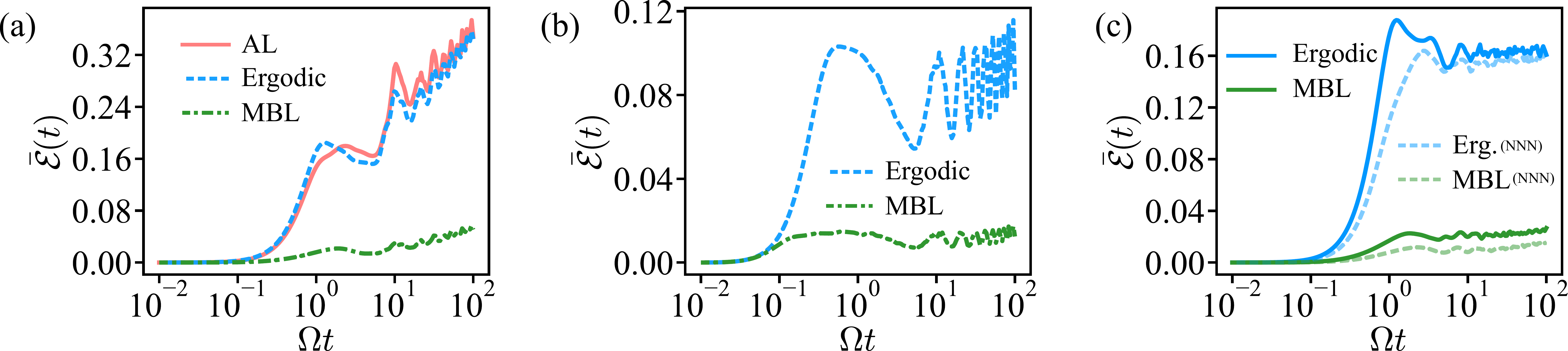}
	\caption{(a,b) The dynamics of average ergotropy with periodic charging field for (a) Ising chain (b) Chimera graph. The number of realizations is taken as $N_{\mathrm{r}}=100$, $\Omega=J_0$ and $\omega^{p}=0.3 J_{0}$. (c) The dynamics of average ergotropy for the various phases of the Ising model compared to the new results with random NNN interactions. We adopt the same parameters as in Fig.~\ref{erg}.}
	\label{periodic}
\end{figure*}

To compute the energy cost we define here the quantity
\begin{align}
	\bar{\Ccal}_{\mathrm{ch}} = \frac{1}{N_{\mathrm{r}}} \sum _{n=1}^{N_{\mathrm{r}}} \frac{\mathcal C_{\mathrm{ch},n}}{\Ecal_{n}^{\mathrm{max}}} ,
\end{align}
where $\mathcal C_{\mathrm{ch},n}$ denotes the charging cost for each model, as given by Eqs.~\eqref{hsnormlinear},~\eqref{hsnormlinearChi},~\eqref{normchargising} and~\eqref{normchargchimera}, in the $n$-th realization. It quantifies the average normalized energy cost of the charging (either with interactions or local fields) as a multiple of the maximum capability $\Ecal_{n}^{\mathrm{max}}$ of the QB for each model. 
We apply this definition to all models considered in this work in order to compare the energy cost for charging by using time-dependent intracell interactions and the cost for the case where we charge the battery by using local fields.
In Fig.~\ref{cost-dis} we present the cost for each model as function of the degree of disorder. For all models considered here as energy storing quantum devices, the charging by local fields is energetically more efficient than the time-dependent interaction engineering, particularly when disorder is strong enough. In particular, the advantage of the local charging comes from the fact that, for high disorder, the interaction dominates over the local charging field as $\mathcal N (H) \rightarrow \mathcal N (H_{\mathrm{ref}})$, providing $\mathcal C_{\mathrm{ch},n} \ll \Ecal_{n}^{\mathrm{max}}$ for each model. We stress that while this limit is consistent, it does not imply that we have zero cost to implement the charging.

\section{Oscillating charging fields} \label{Ap:CharEff}

Let us consider now a charging protocol given by an oscillating classical field acting locally on each quantum cell of the battery. The corresponding charging Hamiltonian is given by 
\begin{align}
	H_{\mathrm{ch}}= \sum_{k}^{N}{\Omega_{k} \cos(\omega^{p}_{k} t) \sigma_{k}^{x}}, \label{Ap:Eq:CharginHamiltonian}
\end{align}
where $\Omega_{k} = \Omega$ and $\omega^{p}_{k}$ denote the amplitude of the periodic charging field and its frequency applied to the \textit{k}-th cell, respectively. First of all, notice that, for such a field, 
the condition $\delta= 6$, such that $\delta J_{0}=6\Omega$, for the MBL phase of the Chimera graph is broken down. Indeed, notice that we have now the change $\Omega \rightarrow \Omega(t) = \Omega \cos(\omega^{p}_{k} t)$. Then, $\delta J_{0}=6\Omega$ is only satisfied as $\omega^{p}_{k} t = 2n\pi$ ($n \in \Zmath$). Therefore, the localization pattern may be affected by the choice of the charging Hamiltonian. By modifying the charging term, we may significantly affect the nature of the models and their quantum phases. 

Now for completeness, let us discuss the battery performance by keeping the phase pattern while slightly replacing Eq.~\eqref{Eq:CharginHamiltonian} for a periodic charging Hamiltonian in Eq.~\eqref{Ap:Eq:CharginHamiltonian}. For simplicity, we set them to be identical for each cell, i.e., $\omega^{p}_{k}=\omega^{p}$. By taking this selection of charging fields, we plot the average ergotropy of the QB [by making use of Eq.~\eqref{eq6}]. 
This is shown in Fig.~\ref{periodic}, where it can be seen that we have the same charging pattern as before. 
Once again, we see that ergotropy storage is small in the MBL phase. Moreover, for the Ising chain, there is a similar behavior for ergodic and AL phases (again with slightly better performance in favor of the AL phase).

\section{Next-to-nearest neighbor disorder effects} \label{Ap:NNN}

In this section, we address the effect of disorder in the next-to-nearest neighbor (NNN) interaction terms. To tackle this point, we randomize the NNN coupling strength, such that $J_{2}$ is randomly selected from the uniform distribution $[-\delta,\delta]$. The disorder strength $\delta$ is set to be the same as for the nearest-neighbor interactions. First of all, we remark that in this scenario $J_{2} \neq 0$ leads to absence of Anderson localized phase in the Ising model, so here we exclude this phase from our analysis. In Fig.~\ref{periodic}{\color{blue}c}, we compare the average ergotropy of the Ising model in its various phases along with the curves for the ergodic and MBL phases with random NNN interactions . As a  main result of this analysis, we observe that this extra randomness reduces the quality of the charging process, since the overall amount of ergotropy is decreased compared to the previous cases where $J_{2}$ is fixed.


\begin{thebibliography}{87}%
	\makeatletter
	\providecommand \@ifxundefined [1]{%
		\@ifx{#1\undefined}
	}%
	\providecommand \@ifnum [1]{%
		\ifnum #1\expandafter \@firstoftwo
		\else \expandafter \@secondoftwo
		\fi
	}%
	\providecommand \@ifx [1]{%
		\ifx #1\expandafter \@firstoftwo
		\else \expandafter \@secondoftwo
		\fi
	}%
	\providecommand \natexlab [1]{#1}%
	\providecommand \enquote  [1]{``#1''}%
	\providecommand \bibnamefont  [1]{#1}%
	\providecommand \bibfnamefont [1]{#1}%
	\providecommand \citenamefont [1]{#1}%
	\providecommand \href@noop [0]{\@secondoftwo}%
	\providecommand \href [0]{\begingroup \@sanitize@url \@href}%
	\providecommand \@href[1]{\@@startlink{#1}\@@href}%
	\providecommand \@@href[1]{\endgroup#1\@@endlink}%
	\providecommand \@sanitize@url [0]{\catcode `\\12\catcode `\$12\catcode
		`\&12\catcode `\#12\catcode `\^12\catcode `\_12\catcode `\%12\relax}%
	\providecommand \@@startlink[1]{}%
	\providecommand \@@endlink[0]{}%
	\providecommand \url  [0]{\begingroup\@sanitize@url \@url }%
	\providecommand \@url [1]{\endgroup\@href {#1}{\urlprefix }}%
	\providecommand \urlprefix  [0]{URL }%
	\providecommand \Eprint [0]{\href }%
	\providecommand \doibase [0]{https://doi.org/}%
	\providecommand \selectlanguage [0]{\@gobble}%
	\providecommand \bibinfo  [0]{\@secondoftwo}%
	\providecommand \bibfield  [0]{\@secondoftwo}%
	\providecommand \translation [1]{[#1]}%
	\providecommand \BibitemOpen [0]{}%
	\providecommand \bibitemStop [0]{}%
	\providecommand \bibitemNoStop [0]{.\EOS\space}%
	\providecommand \EOS [0]{\spacefactor3000\relax}%
	\providecommand \BibitemShut  [1]{\csname bibitem#1\endcsname}%
	\let\auto@bib@innerbib\@empty
	\bibitem [{\citenamefont {Degen}\ \emph {et~al.}(2017)\citenamefont {Degen},
		\citenamefont {Reinhard},\ and\ \citenamefont
		{Cappellaro}}]{degen2017quantum}%
	\BibitemOpen
	\bibfield  {author} {\bibinfo {author} {\bibfnamefont {C.~L.}\ \bibnamefont
			{Degen}}, \bibinfo {author} {\bibfnamefont {F.}~\bibnamefont {Reinhard}},\
		and\ \bibinfo {author} {\bibfnamefont {P.}~\bibnamefont {Cappellaro}},\
	}\bibfield  {title} {\bibinfo {title} {Quantum sensing},\ }\href
	{https://doi.org/10.1103/RevModPhys.89.035002} {\bibfield  {journal}
		{\bibinfo  {journal} {Reviews of modern physics}\ }\textbf {\bibinfo {volume}
			{89}},\ \bibinfo {pages} {035002} (\bibinfo {year} {2017})}\BibitemShut
	{NoStop}%
	\bibitem [{\citenamefont {Zhang}\ and\ \citenamefont
		{Zhuang}(2021)}]{zhang2021distributed}%
	\BibitemOpen
	\bibfield  {author} {\bibinfo {author} {\bibfnamefont {Z.}~\bibnamefont
			{Zhang}}\ and\ \bibinfo {author} {\bibfnamefont {Q.}~\bibnamefont {Zhuang}},\
	}\bibfield  {title} {\bibinfo {title} {Distributed quantum sensing},\ }\href
	{https://doi.org/10.1088/2058-9565/abd4c3} {\bibfield  {journal} {\bibinfo
			{journal} {Quantum Science and Technology}\ }\textbf {\bibinfo {volume}
			{6}},\ \bibinfo {pages} {043001} (\bibinfo {year} {2021})}\BibitemShut
	{NoStop}%
	\bibitem [{\citenamefont {Crawford}\ \emph {et~al.}(2021)\citenamefont
		{Crawford}, \citenamefont {Shugayev}, \citenamefont {Paudel}, \citenamefont
		{Lu}, \citenamefont {Syamlal}, \citenamefont {Ohodnicki}, \citenamefont
		{Chorpening}, \citenamefont {Gentry},\ and\ \citenamefont
		{Duan}}]{crawford2021quantum}%
	\BibitemOpen
	\bibfield  {author} {\bibinfo {author} {\bibfnamefont {S.~E.}\ \bibnamefont
			{Crawford}}, \bibinfo {author} {\bibfnamefont {R.~A.}\ \bibnamefont
			{Shugayev}}, \bibinfo {author} {\bibfnamefont {H.~P.}\ \bibnamefont
			{Paudel}}, \bibinfo {author} {\bibfnamefont {P.}~\bibnamefont {Lu}}, \bibinfo
		{author} {\bibfnamefont {M.}~\bibnamefont {Syamlal}}, \bibinfo {author}
		{\bibfnamefont {P.~R.}\ \bibnamefont {Ohodnicki}}, \bibinfo {author}
		{\bibfnamefont {B.}~\bibnamefont {Chorpening}}, \bibinfo {author}
		{\bibfnamefont {R.}~\bibnamefont {Gentry}},\ and\ \bibinfo {author}
		{\bibfnamefont {Y.}~\bibnamefont {Duan}},\ }\bibfield  {title} {\bibinfo
		{title} {Quantum sensing for energy applications: Review and perspective},\
	}\href {https://doi.org/10.1002/qute.202100049} {\bibfield  {journal}
		{\bibinfo  {journal} {Advanced Quantum Technologies}\ }\textbf {\bibinfo
			{volume} {4}},\ \bibinfo {pages} {2100049} (\bibinfo {year}
		{2021})}\BibitemShut {NoStop}%
	\bibitem [{\citenamefont {Feng}(2019)}]{feng2019review}%
	\BibitemOpen
	\bibfield  {author} {\bibinfo {author} {\bibfnamefont {D.}~\bibnamefont
			{Feng}},\ }\bibfield  {title} {\bibinfo {title} {Review of quantum
			navigation},\ }in\ \href@noop {} {\emph {\bibinfo {booktitle} {IOP Conference
				Series: Earth and Environmental Science}}},\ Vol.\ \bibinfo {volume} {237}\
	(\bibinfo {organization} {IOP Publishing},\ \bibinfo {year} {2019})\ p.\
	\bibinfo {pages} {032027}\BibitemShut {NoStop}%
	\bibitem [{\citenamefont {Poulsen}\ \emph {et~al.}(2022)\citenamefont
		{Poulsen}, \citenamefont {Santos},\ and\ \citenamefont
		{Zinner}}]{PhysRevLett.128.240401}%
	\BibitemOpen
	\bibfield  {author} {\bibinfo {author} {\bibfnamefont {K.}~\bibnamefont
			{Poulsen}}, \bibinfo {author} {\bibfnamefont {A.~C.}\ \bibnamefont
			{Santos}},\ and\ \bibinfo {author} {\bibfnamefont {N.~T.}\ \bibnamefont
			{Zinner}},\ }\bibfield  {title} {\bibinfo {title} {Quantum wheatstone
			bridge},\ }\href {https://doi.org/10.1103/PhysRevLett.128.240401} {\bibfield
		{journal} {\bibinfo  {journal} {Phys. Rev. Lett.}\ }\textbf {\bibinfo
			{volume} {128}},\ \bibinfo {pages} {240401} (\bibinfo {year}
		{2022})}\BibitemShut {NoStop}%
	\bibitem [{\citenamefont {Giovannetti}\ \emph {et~al.}(2011)\citenamefont
		{Giovannetti}, \citenamefont {Lloyd},\ and\ \citenamefont
		{Maccone}}]{giovannetti2011advances}%
	\BibitemOpen
	\bibfield  {author} {\bibinfo {author} {\bibfnamefont {V.}~\bibnamefont
			{Giovannetti}}, \bibinfo {author} {\bibfnamefont {S.}~\bibnamefont {Lloyd}},\
		and\ \bibinfo {author} {\bibfnamefont {L.}~\bibnamefont {Maccone}},\
	}\bibfield  {title} {\bibinfo {title} {Advances in quantum metrology},\
	}\href {https://doi.org/https://doi.org/10.1038/nphoton.2011.35} {\bibfield
		{journal} {\bibinfo  {journal} {Nature photonics}\ }\textbf {\bibinfo
			{volume} {5}},\ \bibinfo {pages} {222} (\bibinfo {year} {2011})}\BibitemShut
	{NoStop}%
	\bibitem [{\citenamefont {Giovannetti}\ \emph {et~al.}(2006)\citenamefont
		{Giovannetti}, \citenamefont {Lloyd},\ and\ \citenamefont
		{Maccone}}]{PhysRevLett.96.010401}%
	\BibitemOpen
	\bibfield  {author} {\bibinfo {author} {\bibfnamefont {V.}~\bibnamefont
			{Giovannetti}}, \bibinfo {author} {\bibfnamefont {S.}~\bibnamefont {Lloyd}},\
		and\ \bibinfo {author} {\bibfnamefont {L.}~\bibnamefont {Maccone}},\
	}\bibfield  {title} {\bibinfo {title} {Quantum metrology},\ }\href
	{https://doi.org/10.1103/PhysRevLett.96.010401} {\bibfield  {journal}
		{\bibinfo  {journal} {Phys. Rev. Lett.}\ }\textbf {\bibinfo {volume} {96}},\
		\bibinfo {pages} {010401} (\bibinfo {year} {2006})}\BibitemShut {NoStop}%
	\bibitem [{\citenamefont {Bennett}\ \emph {et~al.}(1993)\citenamefont
		{Bennett}, \citenamefont {Brassard}, \citenamefont {Cr\'epeau}, \citenamefont
		{Jozsa}, \citenamefont {Peres},\ and\ \citenamefont {Wootters}}]{Bennett:93}%
	\BibitemOpen
	\bibfield  {author} {\bibinfo {author} {\bibfnamefont {C.~H.}\ \bibnamefont
			{Bennett}}, \bibinfo {author} {\bibfnamefont {G.}~\bibnamefont {Brassard}},
		\bibinfo {author} {\bibfnamefont {C.}~\bibnamefont {Cr\'epeau}}, \bibinfo
		{author} {\bibfnamefont {R.}~\bibnamefont {Jozsa}}, \bibinfo {author}
		{\bibfnamefont {A.}~\bibnamefont {Peres}},\ and\ \bibinfo {author}
		{\bibfnamefont {W.~K.}\ \bibnamefont {Wootters}},\ }\bibfield  {title}
	{\bibinfo {title} {Teleporting an unknown quantum state via dual classical
			and einstein-podolsky-rosen channels},\ }\href
	{https://doi.org/10.1103/PhysRevLett.70.1895} {\bibfield  {journal} {\bibinfo
			{journal} {Phys. Rev. Lett.}\ }\textbf {\bibinfo {volume} {70}},\ \bibinfo
		{pages} {1895} (\bibinfo {year} {1993})}\BibitemShut {NoStop}%
	\bibitem [{\citenamefont {Yin}\ \emph {et~al.}(2012)\citenamefont {Yin},
		\citenamefont {Ren}, \citenamefont {Lu}, \citenamefont {Cao}, \citenamefont
		{Yong}, \citenamefont {Wu}, \citenamefont {Liu}, \citenamefont {Liao},
		\citenamefont {Zhou}, \citenamefont {Jiang} \emph {et~al.}}]{yin2012quantum}%
	\BibitemOpen
	\bibfield  {author} {\bibinfo {author} {\bibfnamefont {J.}~\bibnamefont
			{Yin}}, \bibinfo {author} {\bibfnamefont {J.-G.}\ \bibnamefont {Ren}},
		\bibinfo {author} {\bibfnamefont {H.}~\bibnamefont {Lu}}, \bibinfo {author}
		{\bibfnamefont {Y.}~\bibnamefont {Cao}}, \bibinfo {author} {\bibfnamefont
			{H.-L.}\ \bibnamefont {Yong}}, \bibinfo {author} {\bibfnamefont {Y.-P.}\
			\bibnamefont {Wu}}, \bibinfo {author} {\bibfnamefont {C.}~\bibnamefont
			{Liu}}, \bibinfo {author} {\bibfnamefont {S.-K.}\ \bibnamefont {Liao}},
		\bibinfo {author} {\bibfnamefont {F.}~\bibnamefont {Zhou}}, \bibinfo {author}
		{\bibfnamefont {Y.}~\bibnamefont {Jiang}}, \emph {et~al.},\ }\bibfield
	{title} {\bibinfo {title} {Quantum teleportation and entanglement
			distribution over 100-kilometre free-space channels},\ }\href
	{https://doi.org/https://doi.org/10.1038/nature11332} {\bibfield  {journal}
		{\bibinfo  {journal} {Nature}\ }\textbf {\bibinfo {volume} {488}},\ \bibinfo
		{pages} {185} (\bibinfo {year} {2012})}\BibitemShut {NoStop}%
	\bibitem [{\citenamefont {Ma}\ \emph {et~al.}(2012)\citenamefont {Ma},
		\citenamefont {Herbst}, \citenamefont {Scheidl}, \citenamefont {Wang},
		\citenamefont {Kropatschek}, \citenamefont {Naylor}, \citenamefont
		{Wittmann}, \citenamefont {Mech}, \citenamefont {Kofler}, \citenamefont
		{Anisimova} \emph {et~al.}}]{ma2012quantum}%
	\BibitemOpen
	\bibfield  {author} {\bibinfo {author} {\bibfnamefont {X.-S.}\ \bibnamefont
			{Ma}}, \bibinfo {author} {\bibfnamefont {T.}~\bibnamefont {Herbst}}, \bibinfo
		{author} {\bibfnamefont {T.}~\bibnamefont {Scheidl}}, \bibinfo {author}
		{\bibfnamefont {D.}~\bibnamefont {Wang}}, \bibinfo {author} {\bibfnamefont
			{S.}~\bibnamefont {Kropatschek}}, \bibinfo {author} {\bibfnamefont
			{W.}~\bibnamefont {Naylor}}, \bibinfo {author} {\bibfnamefont
			{B.}~\bibnamefont {Wittmann}}, \bibinfo {author} {\bibfnamefont
			{A.}~\bibnamefont {Mech}}, \bibinfo {author} {\bibfnamefont {J.}~\bibnamefont
			{Kofler}}, \bibinfo {author} {\bibfnamefont {E.}~\bibnamefont {Anisimova}},
		\emph {et~al.},\ }\bibfield  {title} {\bibinfo {title} {Quantum teleportation
			over 143 kilometres using active feed-forward},\ }\href
	{https://doi.org/https://doi.org/10.1038/nature11472} {\bibfield  {journal}
		{\bibinfo  {journal} {Nature}\ }\textbf {\bibinfo {volume} {489}},\ \bibinfo
		{pages} {269} (\bibinfo {year} {2012})}\BibitemShut {NoStop}%
	\bibitem [{\citenamefont {{Ren}}\ \emph {et~al.}(2017)\citenamefont {{Ren}},
		\citenamefont {{Xu}}, \citenamefont {{Yong}}, \citenamefont {{Zhang}},
		\citenamefont {{Liao}}, \citenamefont {{Yin}}, \citenamefont {{Liu}},
		\citenamefont {{Cai}}, \citenamefont {{Yang}}, \citenamefont {{Li}},
		\citenamefont {{Yang}}, \citenamefont {{Han}}, \citenamefont {{Yao}},
		\citenamefont {{Li}}, \citenamefont {{Wu}}, \citenamefont {{Wan}},
		\citenamefont {{Liu}}, \citenamefont {{Liu}}, \citenamefont {{Kuang}},
		\citenamefont {{He}}, \citenamefont {{Shang}}, \citenamefont {{Guo}},
		\citenamefont {{Zheng}}, \citenamefont {{Tian}}, \citenamefont {{Zhu}},
		\citenamefont {{Liu}}, \citenamefont {{Lu}}, \citenamefont {{Shu}},
		\citenamefont {{Chen}}, \citenamefont {{Peng}}, \citenamefont {{Wang}},\ and\
		\citenamefont {{Pan}}}]{Ren:17}%
	\BibitemOpen
	\bibfield  {author} {\bibinfo {author} {\bibfnamefont {J.-G.}\ \bibnamefont
			{{Ren}}}, \bibinfo {author} {\bibfnamefont {P.}~\bibnamefont {{Xu}}},
		\bibinfo {author} {\bibfnamefont {H.-L.}\ \bibnamefont {{Yong}}}, \bibinfo
		{author} {\bibfnamefont {L.}~\bibnamefont {{Zhang}}}, \bibinfo {author}
		{\bibfnamefont {S.-K.}\ \bibnamefont {{Liao}}}, \bibinfo {author}
		{\bibfnamefont {J.}~\bibnamefont {{Yin}}}, \bibinfo {author} {\bibfnamefont
			{W.-Y.}\ \bibnamefont {{Liu}}}, \bibinfo {author} {\bibfnamefont {W.-Q.}\
			\bibnamefont {{Cai}}}, \bibinfo {author} {\bibfnamefont {M.}~\bibnamefont
			{{Yang}}}, \bibinfo {author} {\bibfnamefont {L.}~\bibnamefont {{Li}}},
		\bibinfo {author} {\bibfnamefont {K.-X.}\ \bibnamefont {{Yang}}}, \bibinfo
		{author} {\bibfnamefont {X.}~\bibnamefont {{Han}}}, \bibinfo {author}
		{\bibfnamefont {Y.-Q.}\ \bibnamefont {{Yao}}}, \bibinfo {author}
		{\bibfnamefont {J.}~\bibnamefont {{Li}}}, \bibinfo {author} {\bibfnamefont
			{H.-Y.}\ \bibnamefont {{Wu}}}, \bibinfo {author} {\bibfnamefont
			{S.}~\bibnamefont {{Wan}}}, \bibinfo {author} {\bibfnamefont
			{L.}~\bibnamefont {{Liu}}}, \bibinfo {author} {\bibfnamefont {D.-Q.}\
			\bibnamefont {{Liu}}}, \bibinfo {author} {\bibfnamefont {Y.-W.}\ \bibnamefont
			{{Kuang}}}, \bibinfo {author} {\bibfnamefont {Z.-P.}\ \bibnamefont {{He}}},
		\bibinfo {author} {\bibfnamefont {P.}~\bibnamefont {{Shang}}}, \bibinfo
		{author} {\bibfnamefont {C.}~\bibnamefont {{Guo}}}, \bibinfo {author}
		{\bibfnamefont {R.-H.}\ \bibnamefont {{Zheng}}}, \bibinfo {author}
		{\bibfnamefont {K.}~\bibnamefont {{Tian}}}, \bibinfo {author} {\bibfnamefont
			{Z.-C.}\ \bibnamefont {{Zhu}}}, \bibinfo {author} {\bibfnamefont {N.-L.}\
			\bibnamefont {{Liu}}}, \bibinfo {author} {\bibfnamefont {C.-Y.}\ \bibnamefont
			{{Lu}}}, \bibinfo {author} {\bibfnamefont {R.}~\bibnamefont {{Shu}}},
		\bibinfo {author} {\bibfnamefont {Y.-A.}\ \bibnamefont {{Chen}}}, \bibinfo
		{author} {\bibfnamefont {C.-Z.}\ \bibnamefont {{Peng}}}, \bibinfo {author}
		{\bibfnamefont {J.-Y.}\ \bibnamefont {{Wang}}},\ and\ \bibinfo {author}
		{\bibfnamefont {J.-W.}\ \bibnamefont {{Pan}}},\ }\bibfield  {title} {\bibinfo
		{title} {Ground-to-satellite quantum teleportation},\ }\href
	{https://doi.org/10.1038/nature23675} {\bibfield  {journal} {\bibinfo
			{journal} {Nature}\ }\textbf {\bibinfo {volume} {549}},\ \bibinfo {pages}
		{70} (\bibinfo {year} {2017})}\BibitemShut {NoStop}%
	\bibitem [{\citenamefont {Geppert}(2000)}]{geppert2000quantum}%
	\BibitemOpen
	\bibfield  {author} {\bibinfo {author} {\bibfnamefont {L.}~\bibnamefont
			{Geppert}},\ }\bibfield  {title} {\bibinfo {title} {Quantum transistors:
			toward nanoelectronics},\ }\href {https://doi.org/10.1109/6.866283}
	{\bibfield  {journal} {\bibinfo  {journal} {IEEE Spectr.}\ }\textbf {\bibinfo
			{volume} {37}},\ \bibinfo {pages} {46} (\bibinfo {year} {2000})}\BibitemShut
	{NoStop}%
	\bibitem [{\citenamefont {Marchukov}\ \emph {et~al.}(2016)\citenamefont
		{Marchukov}, \citenamefont {Volosniev}, \citenamefont {Valiente},
		\citenamefont {Petrosyan},\ and\ \citenamefont {Zinner}}]{Marchukov:16}%
	\BibitemOpen
	\bibfield  {author} {\bibinfo {author} {\bibfnamefont {O.~V.}\ \bibnamefont
			{Marchukov}}, \bibinfo {author} {\bibfnamefont {A.~G.}\ \bibnamefont
			{Volosniev}}, \bibinfo {author} {\bibfnamefont {M.}~\bibnamefont {Valiente}},
		\bibinfo {author} {\bibfnamefont {D.}~\bibnamefont {Petrosyan}},\ and\
		\bibinfo {author} {\bibfnamefont {N.}~\bibnamefont {Zinner}},\ }\bibfield
	{title} {\bibinfo {title} {Quantum spin transistor with a heisenberg spin
			chain},\ }\href {https://doi.org/https://doi.org/10.1038/ncomms13070}
	{\bibfield  {journal} {\bibinfo  {journal} {Nature communications}\ }\textbf
		{\bibinfo {volume} {7}},\ \bibinfo {pages} {1} (\bibinfo {year}
		{2016})}\BibitemShut {NoStop}%
	\bibitem [{\citenamefont {Joulain}\ \emph {et~al.}(2016)\citenamefont
		{Joulain}, \citenamefont {Drevillon}, \citenamefont {Ezzahri},\ and\
		\citenamefont {Ordonez-Miranda}}]{joulain2016quantum}%
	\BibitemOpen
	\bibfield  {author} {\bibinfo {author} {\bibfnamefont {K.}~\bibnamefont
			{Joulain}}, \bibinfo {author} {\bibfnamefont {J.}~\bibnamefont {Drevillon}},
		\bibinfo {author} {\bibfnamefont {Y.}~\bibnamefont {Ezzahri}},\ and\ \bibinfo
		{author} {\bibfnamefont {J.}~\bibnamefont {Ordonez-Miranda}},\ }\bibfield
	{title} {\bibinfo {title} {Quantum thermal transistor},\ }\href
	{https://doi.org/10.1103/PhysRevLett.116.200601} {\bibfield  {journal}
		{\bibinfo  {journal} {Phys. Rev. Lett.}\ }\textbf {\bibinfo {volume} {116}},\
		\bibinfo {pages} {200601} (\bibinfo {year} {2016})}\BibitemShut {NoStop}%
	\bibitem [{\citenamefont {{Loft}}\ \emph {et~al.}(2018)\citenamefont {{Loft}},
		\citenamefont {{Kristensen}}, \citenamefont {{Andersen}},\ and\ \citenamefont
		{{Zinner}}}]{Loft:18}%
	\BibitemOpen
	\bibfield  {author} {\bibinfo {author} {\bibfnamefont {N.~J.~S.}\
			\bibnamefont {{Loft}}}, \bibinfo {author} {\bibfnamefont {L.~B.}\
			\bibnamefont {{Kristensen}}}, \bibinfo {author} {\bibfnamefont {C.~K.}\
			\bibnamefont {{Andersen}}},\ and\ \bibinfo {author} {\bibfnamefont {N.~T.}\
			\bibnamefont {{Zinner}}},\ }\bibfield  {title} {\bibinfo {title} {{Quantum
				spin transistors in superconducting circuits}},\ }\href@noop {} {\bibfield
		{journal} {\bibinfo  {journal} {arXiv e-prints}\ ,\ \bibinfo {eid}
			{arXiv:1802.04292}} (\bibinfo {year} {2018})},\ \Eprint
	{https://arxiv.org/abs/1802.04292} {arXiv:1802.04292 [quant-ph]} \BibitemShut
	{NoStop}%
	\bibitem [{\citenamefont {{de Ponte}}\ and\ \citenamefont
		{{Santos}}(2019)}]{dePonte:19}%
	\BibitemOpen
	\bibfield  {author} {\bibinfo {author} {\bibfnamefont {M.~A.}\ \bibnamefont
			{{de Ponte}}}\ and\ \bibinfo {author} {\bibfnamefont {A.~C.}\ \bibnamefont
			{{Santos}}},\ }\bibfield  {title} {\bibinfo {title} {Shortening time scale to
			reduce thermal effects in quantum transistors},\ }\href
	{https://doi.org/10.1038/s41598-019-46902-5} {\bibfield  {journal} {\bibinfo
			{journal} {Sci. Rep.}\ }\textbf {\bibinfo {volume} {9}},\ \bibinfo {pages}
		{10470} (\bibinfo {year} {2019})}\BibitemShut {NoStop}%
	\bibitem [{\citenamefont {McRae}\ \emph {et~al.}(2019)\citenamefont {McRae},
		\citenamefont {Wei},\ and\ \citenamefont {Champagne}}]{mcrae2019graphene}%
	\BibitemOpen
	\bibfield  {author} {\bibinfo {author} {\bibfnamefont {A.}~\bibnamefont
			{McRae}}, \bibinfo {author} {\bibfnamefont {G.}~\bibnamefont {Wei}},\ and\
		\bibinfo {author} {\bibfnamefont {A.}~\bibnamefont {Champagne}},\ }\bibfield
	{title} {\bibinfo {title} {Graphene quantum strain transistors},\ }\href
	{https://doi.org/10.1103/PhysRevApplied.11.054019} {\bibfield  {journal}
		{\bibinfo  {journal} {Phys. Rev. Appl.}\ }\textbf {\bibinfo {volume} {11}},\
		\bibinfo {pages} {054019} (\bibinfo {year} {2019})}\BibitemShut {NoStop}%
	\bibitem [{\citenamefont {Shukla}\ and\ \citenamefont
		{Eliasson}(2008)}]{shukla2008nonlinear}%
	\BibitemOpen
	\bibfield  {author} {\bibinfo {author} {\bibfnamefont {P.~K.}\ \bibnamefont
			{Shukla}}\ and\ \bibinfo {author} {\bibfnamefont {B.}~\bibnamefont
			{Eliasson}},\ }\bibfield  {title} {\bibinfo {title} {Nonlinear theory for a
			quantum diode in a dense fermi magnetoplasma},\ }\href
	{https://doi.org/10.1103/PhysRevLett.100.036801} {\bibfield  {journal}
		{\bibinfo  {journal} {Phys. Rev. Lett.}\ }\textbf {\bibinfo {volume} {100}},\
		\bibinfo {pages} {036801} (\bibinfo {year} {2008})}\BibitemShut {NoStop}%
	\bibitem [{\citenamefont {Shen}\ \emph {et~al.}(2014)\citenamefont {Shen},
		\citenamefont {Zhou},\ and\ \citenamefont {Yi}}]{shen2014quantum}%
	\BibitemOpen
	\bibfield  {author} {\bibinfo {author} {\bibfnamefont {H.}~\bibnamefont
			{Shen}}, \bibinfo {author} {\bibfnamefont {Y.}~\bibnamefont {Zhou}},\ and\
		\bibinfo {author} {\bibfnamefont {X.}~\bibnamefont {Yi}},\ }\bibfield
	{title} {\bibinfo {title} {Quantum optical diode with semiconductor
			microcavities},\ }\href {https://doi.org/10.1103/PhysRevA.90.023849}
	{\bibfield  {journal} {\bibinfo  {journal} {Phys. Rev. A}\ }\textbf {\bibinfo
			{volume} {90}},\ \bibinfo {pages} {023849} (\bibinfo {year}
		{2014})}\BibitemShut {NoStop}%
	\bibitem [{\citenamefont {Devoret}\ and\ \citenamefont
		{Glattli}(1998)}]{Devoret:98}%
	\BibitemOpen
	\bibfield  {author} {\bibinfo {author} {\bibfnamefont {M.~H.}\ \bibnamefont
			{Devoret}}\ and\ \bibinfo {author} {\bibfnamefont {C.}~\bibnamefont
			{Glattli}},\ }\bibfield  {title} {\bibinfo {title} {Single-electron
			transistors},\ }\href@noop {} {\bibfield  {journal} {\bibinfo  {journal}
			{Phys. World}\ }\textbf {\bibinfo {volume} {11}},\ \bibinfo {pages} {29}
		(\bibinfo {year} {1998})}\BibitemShut {NoStop}%
	\bibitem [{\citenamefont {Arute}\ \emph {et~al.}(2019)\citenamefont {Arute},
		\citenamefont {Arya}, \citenamefont {Babbush}, \citenamefont {Bacon},
		\citenamefont {Bardin}, \citenamefont {Barends}, \citenamefont {Biswas},
		\citenamefont {Boixo}, \citenamefont {Brandao}, \citenamefont {Buell},
		\citenamefont {Burkett}, \citenamefont {Chen}, \citenamefont {Chen},
		\citenamefont {Chiaro}, \citenamefont {Collins}, \citenamefont {Courtney},
		\citenamefont {Dunsworth}, \citenamefont {Farhi}, \citenamefont {Foxen},
		\citenamefont {Fowler}, \citenamefont {Gidney}, \citenamefont {Giustina},
		\citenamefont {Graff}, \citenamefont {Guerin}, \citenamefont {Habegger},
		\citenamefont {Harrigan}, \citenamefont {Hartmann}, \citenamefont {Ho},
		\citenamefont {Hoffmann}, \citenamefont {Huang}, \citenamefont {Humble},
		\citenamefont {Isakov}, \citenamefont {Jeffrey}, \citenamefont {Jiang},
		\citenamefont {Kafri}, \citenamefont {Kechedzhi}, \citenamefont {Kelly},
		\citenamefont {Klimov}, \citenamefont {Knysh}, \citenamefont {Korotkov},
		\citenamefont {Kostritsa}, \citenamefont {Landhuis}, \citenamefont
		{Lindmark}, \citenamefont {Lucero}, \citenamefont {Lyakh}, \citenamefont
		{Mandrà}, \citenamefont {McClean}, \citenamefont {McEwen}, \citenamefont
		{Megrant}, \citenamefont {Mi}, \citenamefont {Michielsen}, \citenamefont
		{Mohseni}, \citenamefont {Mutus}, \citenamefont {Naaman}, \citenamefont
		{Neeley}, \citenamefont {Neill}, \citenamefont {Niu}, \citenamefont {Ostby},
		\citenamefont {Petukhov}, \citenamefont {Platt}, \citenamefont {Quintana},
		\citenamefont {Rieffel}, \citenamefont {Roushan}, \citenamefont {Rubin},
		\citenamefont {Sank}, \citenamefont {Satzinger}, \citenamefont {Smelyanskiy},
		\citenamefont {Sung}, \citenamefont {Trevithick}, \citenamefont
		{Vainsencher}, \citenamefont {Villalonga}, \citenamefont {White},
		\citenamefont {Yao}, \citenamefont {Yeh}, \citenamefont {Zalcman},
		\citenamefont {Neven},\ and\ \citenamefont {Martinis}}]{Arute:19}%
	\BibitemOpen
	\bibfield  {author} {\bibinfo {author} {\bibfnamefont {F.}~\bibnamefont
			{Arute}}, \bibinfo {author} {\bibfnamefont {K.}~\bibnamefont {Arya}},
		\bibinfo {author} {\bibfnamefont {R.}~\bibnamefont {Babbush}}, \bibinfo
		{author} {\bibfnamefont {D.}~\bibnamefont {Bacon}}, \bibinfo {author}
		{\bibfnamefont {J.~C.}\ \bibnamefont {Bardin}}, \bibinfo {author}
		{\bibfnamefont {R.}~\bibnamefont {Barends}}, \bibinfo {author} {\bibfnamefont
			{R.}~\bibnamefont {Biswas}}, \bibinfo {author} {\bibfnamefont
			{S.}~\bibnamefont {Boixo}}, \bibinfo {author} {\bibfnamefont {F.~G. S.~L.}\
			\bibnamefont {Brandao}}, \bibinfo {author} {\bibfnamefont {D.~A.}\
			\bibnamefont {Buell}}, \bibinfo {author} {\bibfnamefont {B.}~\bibnamefont
			{Burkett}}, \bibinfo {author} {\bibfnamefont {Y.}~\bibnamefont {Chen}},
		\bibinfo {author} {\bibfnamefont {Z.}~\bibnamefont {Chen}}, \bibinfo {author}
		{\bibfnamefont {B.}~\bibnamefont {Chiaro}}, \bibinfo {author} {\bibfnamefont
			{R.}~\bibnamefont {Collins}}, \bibinfo {author} {\bibfnamefont
			{W.}~\bibnamefont {Courtney}}, \bibinfo {author} {\bibfnamefont
			{A.}~\bibnamefont {Dunsworth}}, \bibinfo {author} {\bibfnamefont
			{E.}~\bibnamefont {Farhi}}, \bibinfo {author} {\bibfnamefont
			{B.}~\bibnamefont {Foxen}}, \bibinfo {author} {\bibfnamefont
			{A.}~\bibnamefont {Fowler}}, \bibinfo {author} {\bibfnamefont
			{C.}~\bibnamefont {Gidney}}, \bibinfo {author} {\bibfnamefont
			{M.}~\bibnamefont {Giustina}}, \bibinfo {author} {\bibfnamefont
			{R.}~\bibnamefont {Graff}}, \bibinfo {author} {\bibfnamefont
			{K.}~\bibnamefont {Guerin}}, \bibinfo {author} {\bibfnamefont
			{S.}~\bibnamefont {Habegger}}, \bibinfo {author} {\bibfnamefont {M.~P.}\
			\bibnamefont {Harrigan}}, \bibinfo {author} {\bibfnamefont {M.~J.}\
			\bibnamefont {Hartmann}}, \bibinfo {author} {\bibfnamefont {A.}~\bibnamefont
			{Ho}}, \bibinfo {author} {\bibfnamefont {M.}~\bibnamefont {Hoffmann}},
		\bibinfo {author} {\bibfnamefont {T.}~\bibnamefont {Huang}}, \bibinfo
		{author} {\bibfnamefont {T.~S.}\ \bibnamefont {Humble}}, \bibinfo {author}
		{\bibfnamefont {S.~V.}\ \bibnamefont {Isakov}}, \bibinfo {author}
		{\bibfnamefont {E.}~\bibnamefont {Jeffrey}}, \bibinfo {author} {\bibfnamefont
			{Z.}~\bibnamefont {Jiang}}, \bibinfo {author} {\bibfnamefont
			{D.}~\bibnamefont {Kafri}}, \bibinfo {author} {\bibfnamefont
			{K.}~\bibnamefont {Kechedzhi}}, \bibinfo {author} {\bibfnamefont
			{J.}~\bibnamefont {Kelly}}, \bibinfo {author} {\bibfnamefont {P.~V.}\
			\bibnamefont {Klimov}}, \bibinfo {author} {\bibfnamefont {S.}~\bibnamefont
			{Knysh}}, \bibinfo {author} {\bibfnamefont {A.}~\bibnamefont {Korotkov}},
		\bibinfo {author} {\bibfnamefont {F.}~\bibnamefont {Kostritsa}}, \bibinfo
		{author} {\bibfnamefont {D.}~\bibnamefont {Landhuis}}, \bibinfo {author}
		{\bibfnamefont {M.}~\bibnamefont {Lindmark}}, \bibinfo {author}
		{\bibfnamefont {E.}~\bibnamefont {Lucero}}, \bibinfo {author} {\bibfnamefont
			{D.}~\bibnamefont {Lyakh}}, \bibinfo {author} {\bibfnamefont
			{S.}~\bibnamefont {Mandrà}}, \bibinfo {author} {\bibfnamefont {J.~R.}\
			\bibnamefont {McClean}}, \bibinfo {author} {\bibfnamefont {M.}~\bibnamefont
			{McEwen}}, \bibinfo {author} {\bibfnamefont {A.}~\bibnamefont {Megrant}},
		\bibinfo {author} {\bibfnamefont {X.}~\bibnamefont {Mi}}, \bibinfo {author}
		{\bibfnamefont {K.}~\bibnamefont {Michielsen}}, \bibinfo {author}
		{\bibfnamefont {M.}~\bibnamefont {Mohseni}}, \bibinfo {author} {\bibfnamefont
			{J.}~\bibnamefont {Mutus}}, \bibinfo {author} {\bibfnamefont
			{O.}~\bibnamefont {Naaman}}, \bibinfo {author} {\bibfnamefont
			{M.}~\bibnamefont {Neeley}}, \bibinfo {author} {\bibfnamefont
			{C.}~\bibnamefont {Neill}}, \bibinfo {author} {\bibfnamefont {M.~Y.}\
			\bibnamefont {Niu}}, \bibinfo {author} {\bibfnamefont {E.}~\bibnamefont
			{Ostby}}, \bibinfo {author} {\bibfnamefont {A.}~\bibnamefont {Petukhov}},
		\bibinfo {author} {\bibfnamefont {J.~C.}\ \bibnamefont {Platt}}, \bibinfo
		{author} {\bibfnamefont {C.}~\bibnamefont {Quintana}}, \bibinfo {author}
		{\bibfnamefont {E.~G.}\ \bibnamefont {Rieffel}}, \bibinfo {author}
		{\bibfnamefont {P.}~\bibnamefont {Roushan}}, \bibinfo {author} {\bibfnamefont
			{N.~C.}\ \bibnamefont {Rubin}}, \bibinfo {author} {\bibfnamefont
			{D.}~\bibnamefont {Sank}}, \bibinfo {author} {\bibfnamefont {K.~J.}\
			\bibnamefont {Satzinger}}, \bibinfo {author} {\bibfnamefont {V.}~\bibnamefont
			{Smelyanskiy}}, \bibinfo {author} {\bibfnamefont {K.~J.}\ \bibnamefont
			{Sung}}, \bibinfo {author} {\bibfnamefont {M.~D.}\ \bibnamefont
			{Trevithick}}, \bibinfo {author} {\bibfnamefont {A.}~\bibnamefont
			{Vainsencher}}, \bibinfo {author} {\bibfnamefont {B.}~\bibnamefont
			{Villalonga}}, \bibinfo {author} {\bibfnamefont {T.}~\bibnamefont {White}},
		\bibinfo {author} {\bibfnamefont {Z.~J.}\ \bibnamefont {Yao}}, \bibinfo
		{author} {\bibfnamefont {P.}~\bibnamefont {Yeh}}, \bibinfo {author}
		{\bibfnamefont {A.}~\bibnamefont {Zalcman}}, \bibinfo {author} {\bibfnamefont
			{H.}~\bibnamefont {Neven}},\ and\ \bibinfo {author} {\bibfnamefont {J.~M.}\
			\bibnamefont {Martinis}},\ }\bibfield  {title} {\bibinfo {title} {Quantum
			supremacy using a programmable superconducting processor},\ }\href
	{https://doi.org/10.1038/s41586-019-1666-5} {\bibfield  {journal} {\bibinfo
			{journal} {Nature}\ }\textbf {\bibinfo {volume} {574}},\ \bibinfo {pages}
		{505} (\bibinfo {year} {2019})}\BibitemShut {NoStop}%
	\bibitem [{\citenamefont {Bacon}\ and\ \citenamefont
		{Flammia}(2009)}]{Bacon:09}%
	\BibitemOpen
	\bibfield  {author} {\bibinfo {author} {\bibfnamefont {D.}~\bibnamefont
			{Bacon}}\ and\ \bibinfo {author} {\bibfnamefont {S.~T.}\ \bibnamefont
			{Flammia}},\ }\bibfield  {title} {\bibinfo {title} {Adiabatic gate
			teleportation},\ }\href {https://doi.org/10.1103/PhysRevLett.103.120504}
	{\bibfield  {journal} {\bibinfo  {journal} {Phys. Rev. Lett.}\ }\textbf
		{\bibinfo {volume} {103}},\ \bibinfo {pages} {120504} (\bibinfo {year}
		{2009})}\BibitemShut {NoStop}%
	\bibitem [{\citenamefont {Santos}\ \emph {et~al.}(2016)\citenamefont {Santos},
		\citenamefont {Silva},\ and\ \citenamefont {Sarandy}}]{Santos:16}%
	\BibitemOpen
	\bibfield  {author} {\bibinfo {author} {\bibfnamefont {A.~C.}\ \bibnamefont
			{Santos}}, \bibinfo {author} {\bibfnamefont {R.~D.}\ \bibnamefont {Silva}},\
		and\ \bibinfo {author} {\bibfnamefont {M.~S.}\ \bibnamefont {Sarandy}},\
	}\bibfield  {title} {\bibinfo {title} {Shortcut to adiabatic gate
			teleportation},\ }\href {https://doi.org/10.1103/PhysRevA.93.012311}
	{\bibfield  {journal} {\bibinfo  {journal} {Phys. Rev. A}\ }\textbf {\bibinfo
			{volume} {93}},\ \bibinfo {pages} {012311} (\bibinfo {year}
		{2016})}\BibitemShut {NoStop}%
	\bibitem [{\citenamefont {Gottesman}\ and\ \citenamefont
		{Chuang}(1999)}]{Gottesman:99}%
	\BibitemOpen
	\bibfield  {author} {\bibinfo {author} {\bibfnamefont {D.}~\bibnamefont
			{Gottesman}}\ and\ \bibinfo {author} {\bibfnamefont {I.~L.}\ \bibnamefont
			{Chuang}},\ }\bibfield  {title} {\bibinfo {title} {Demonstrating the
			viability of universal quantum computation using teleportation and
			single-qubit operations},\ }\href
	{https://doi.org/https://doi.org/10.1038/46503} {\bibfield  {journal}
		{\bibinfo  {journal} {Nature}\ }\textbf {\bibinfo {volume} {402}},\ \bibinfo
		{pages} {390} (\bibinfo {year} {1999})}\BibitemShut {NoStop}%
	\bibitem [{\citenamefont {Briegel}\ \emph {et~al.}(1998)\citenamefont
		{Briegel}, \citenamefont {D\"ur}, \citenamefont {Cirac},\ and\ \citenamefont
		{Zoller}}]{PhysRevLett.81.5932}%
	\BibitemOpen
	\bibfield  {author} {\bibinfo {author} {\bibfnamefont {H.-J.}\ \bibnamefont
			{Briegel}}, \bibinfo {author} {\bibfnamefont {W.}~\bibnamefont {D\"ur}},
		\bibinfo {author} {\bibfnamefont {J.~I.}\ \bibnamefont {Cirac}},\ and\
		\bibinfo {author} {\bibfnamefont {P.}~\bibnamefont {Zoller}},\ }\bibfield
	{title} {\bibinfo {title} {Quantum repeaters: The role of imperfect local
			operations in quantum communication},\ }\href
	{https://doi.org/10.1103/PhysRevLett.81.5932} {\bibfield  {journal} {\bibinfo
			{journal} {Phys. Rev. Lett.}\ }\textbf {\bibinfo {volume} {81}},\ \bibinfo
		{pages} {5932} (\bibinfo {year} {1998})}\BibitemShut {NoStop}%
	\bibitem [{\citenamefont {DiVincenzo}\ and\ \citenamefont
		{Loss}(1999)}]{divincenzo1999quantum}%
	\BibitemOpen
	\bibfield  {author} {\bibinfo {author} {\bibfnamefont {D.~P.}\ \bibnamefont
			{DiVincenzo}}\ and\ \bibinfo {author} {\bibfnamefont {D.}~\bibnamefont
			{Loss}},\ }\bibfield  {title} {\bibinfo {title} {Quantum computers and
			quantum coherence},\ }\href {https://doi.org/10.1016/S0304-8853(99)00315-7}
	{\bibfield  {journal} {\bibinfo  {journal} {J. Magn. Magn. Mater.}\ }\textbf
		{\bibinfo {volume} {200}},\ \bibinfo {pages} {202} (\bibinfo {year}
		{1999})}\BibitemShut {NoStop}%
	\bibitem [{\citenamefont {Campaioli}\ \emph {et~al.}(2018)\citenamefont
		{Campaioli}, \citenamefont {Pollock},\ and\ \citenamefont
		{Vinjanampathy}}]{campaioli2018quantum}%
	\BibitemOpen
	\bibfield  {author} {\bibinfo {author} {\bibfnamefont {F.}~\bibnamefont
			{Campaioli}}, \bibinfo {author} {\bibfnamefont {F.~A.}\ \bibnamefont
			{Pollock}},\ and\ \bibinfo {author} {\bibfnamefont {S.}~\bibnamefont
			{Vinjanampathy}},\ }\bibfield  {title} {\bibinfo {title} {Quantum
			batteries},\ }in\ \href@noop {} {\emph {\bibinfo {booktitle} {Thermodynamics
				in the Quantum Regime}}}\ (\bibinfo  {publisher} {Springer},\ \bibinfo {year}
	{2018})\ pp.\ \bibinfo {pages} {207--225}\BibitemShut {NoStop}%
	\bibitem [{\citenamefont {Bhattacharjee}\ and\ \citenamefont
		{Dutta}(2021)}]{bhattacharjee2021quantum}%
	\BibitemOpen
	\bibfield  {author} {\bibinfo {author} {\bibfnamefont {S.}~\bibnamefont
			{Bhattacharjee}}\ and\ \bibinfo {author} {\bibfnamefont {A.}~\bibnamefont
			{Dutta}},\ }\bibfield  {title} {\bibinfo {title} {Quantum thermal machines
			and batteries},\ }\href {https://doi.org/10.1140/epjb/s10051-021-00235-3}
	{\bibfield  {journal} {\bibinfo  {journal} {The European Physical Journal B}\
		}\textbf {\bibinfo {volume} {94}},\ \bibinfo {pages} {1} (\bibinfo {year}
		{2021})}\BibitemShut {NoStop}%
	\bibitem [{\citenamefont {Alicki}\ and\ \citenamefont
		{Fannes}(2013)}]{Alicki:13}%
	\BibitemOpen
	\bibfield  {author} {\bibinfo {author} {\bibfnamefont {R.}~\bibnamefont
			{Alicki}}\ and\ \bibinfo {author} {\bibfnamefont {M.}~\bibnamefont
			{Fannes}},\ }\bibfield  {title} {\bibinfo {title} {Entanglement boost for
			extractable work from ensembles of quantum batteries},\ }\href
	{https://doi.org/10.1103/PhysRevE.87.042123} {\bibfield  {journal} {\bibinfo
			{journal} {Phys. Rev. E}\ }\textbf {\bibinfo {volume} {87}},\ \bibinfo
		{pages} {042123} (\bibinfo {year} {2013})}\BibitemShut {NoStop}%
	\bibitem [{\citenamefont {Santos}\ \emph {et~al.}(2019)\citenamefont {Santos},
		\citenamefont {\c{C}akmak}, \citenamefont {Campbell},\ and\ \citenamefont
		{Zinner}}]{Santos:19-a}%
	\BibitemOpen
	\bibfield  {author} {\bibinfo {author} {\bibfnamefont {A.~C.}\ \bibnamefont
			{Santos}}, \bibinfo {author} {\bibfnamefont {B.}~\bibnamefont {\c{C}akmak}},
		\bibinfo {author} {\bibfnamefont {S.}~\bibnamefont {Campbell}},\ and\
		\bibinfo {author} {\bibfnamefont {N.~T.}\ \bibnamefont {Zinner}},\ }\bibfield
	{title} {\bibinfo {title} {Stable adiabatic quantum batteries},\ }\href
	{https://doi.org/10.1103/PhysRevE.100.032107} {\bibfield  {journal} {\bibinfo
			{journal} {Phys. Rev. E}\ }\textbf {\bibinfo {volume} {100}},\ \bibinfo
		{pages} {032107} (\bibinfo {year} {2019})}\BibitemShut {NoStop}%
	\bibitem [{\citenamefont {Andolina}\ \emph
		{et~al.}(2019{\natexlab{a}})\citenamefont {Andolina}, \citenamefont {Keck},
		\citenamefont {Mari}, \citenamefont {Giovannetti},\ and\ \citenamefont
		{Polini}}]{Andolina:19}%
	\BibitemOpen
	\bibfield  {author} {\bibinfo {author} {\bibfnamefont {G.~M.}\ \bibnamefont
			{Andolina}}, \bibinfo {author} {\bibfnamefont {M.}~\bibnamefont {Keck}},
		\bibinfo {author} {\bibfnamefont {A.}~\bibnamefont {Mari}}, \bibinfo {author}
		{\bibfnamefont {V.}~\bibnamefont {Giovannetti}},\ and\ \bibinfo {author}
		{\bibfnamefont {M.}~\bibnamefont {Polini}},\ }\bibfield  {title} {\bibinfo
		{title} {Quantum versus classical many-body batteries},\ }\href
	{https://doi.org/10.1103/PhysRevB.99.205437} {\bibfield  {journal} {\bibinfo
			{journal} {Phys. Rev. B}\ }\textbf {\bibinfo {volume} {99}},\ \bibinfo
		{pages} {205437} (\bibinfo {year} {2019}{\natexlab{a}})}\BibitemShut
	{NoStop}%
	\bibitem [{\citenamefont {Shaghaghi}\ \emph {et~al.}(2022)\citenamefont
		{Shaghaghi}, \citenamefont {Singh}, \citenamefont {Benenti},\ and\
		\citenamefont {Rosa}}]{Shaghaghi:micromaser}%
	\BibitemOpen
	\bibfield  {author} {\bibinfo {author} {\bibfnamefont {V.}~\bibnamefont
			{Shaghaghi}}, \bibinfo {author} {\bibfnamefont {V.}~\bibnamefont {Singh}},
		\bibinfo {author} {\bibfnamefont {G.}~\bibnamefont {Benenti}},\ and\ \bibinfo
		{author} {\bibfnamefont {D.}~\bibnamefont {Rosa}},\ }\bibfield  {title}
	{\bibinfo {title} {Micromasers as quantum batteries},\ }\href
	{https://doi.org/10.1088/2058-9565/ac8829} {\bibfield  {journal} {\bibinfo
			{journal} {Quantum Science and Technology}\ }\textbf {\bibinfo {volume}
			{7}},\ \bibinfo {pages} {04LT01} (\bibinfo {year} {2022})}\BibitemShut
	{NoStop}%
	\bibitem [{\citenamefont {Downing}\ and\ \citenamefont
		{Sturges}(2022)}]{Downing_2022}%
	\BibitemOpen
	\bibfield  {author} {\bibinfo {author} {\bibfnamefont {C.~A.}\ \bibnamefont
			{Downing}}\ and\ \bibinfo {author} {\bibfnamefont {T.~J.}\ \bibnamefont
			{Sturges}},\ }\bibfield  {title} {\bibinfo {title} {Directionality between
			driven-dissipative resonators},\ }\href
	{https://doi.org/10.1209/0295-5075/ac9ad6} {\bibfield  {journal} {\bibinfo
			{journal} {Europhysics Letters}\ }\textbf {\bibinfo {volume} {140}},\
		\bibinfo {pages} {35001} (\bibinfo {year} {2022})}\BibitemShut {NoStop}%
	\bibitem [{\citenamefont {Andolina}\ \emph
		{et~al.}(2019{\natexlab{b}})\citenamefont {Andolina}, \citenamefont {Keck},
		\citenamefont {Mari}, \citenamefont {Campisi}, \citenamefont {Giovannetti},\
		and\ \citenamefont {Polini}}]{PRL_Andolina}%
	\BibitemOpen
	\bibfield  {author} {\bibinfo {author} {\bibfnamefont {G.~M.}\ \bibnamefont
			{Andolina}}, \bibinfo {author} {\bibfnamefont {M.}~\bibnamefont {Keck}},
		\bibinfo {author} {\bibfnamefont {A.}~\bibnamefont {Mari}}, \bibinfo {author}
		{\bibfnamefont {M.}~\bibnamefont {Campisi}}, \bibinfo {author} {\bibfnamefont
			{V.}~\bibnamefont {Giovannetti}},\ and\ \bibinfo {author} {\bibfnamefont
			{M.}~\bibnamefont {Polini}},\ }\bibfield  {title} {\bibinfo {title}
		{Extractable work, the role of correlations, and asymptotic freedom in
			quantum batteries},\ }\href {https://doi.org/10.1103/PhysRevLett.122.047702}
	{\bibfield  {journal} {\bibinfo  {journal} {Phys. Rev. Lett.}\ }\textbf
		{\bibinfo {volume} {122}},\ \bibinfo {pages} {047702} (\bibinfo {year}
		{2019}{\natexlab{b}})}\BibitemShut {NoStop}%
	\bibitem [{\citenamefont {Farina}\ \emph {et~al.}(2019)\citenamefont {Farina},
		\citenamefont {Andolina}, \citenamefont {Mari}, \citenamefont {Polini},\ and\
		\citenamefont {Giovannetti}}]{PRB2019Batteries}%
	\BibitemOpen
	\bibfield  {author} {\bibinfo {author} {\bibfnamefont {D.}~\bibnamefont
			{Farina}}, \bibinfo {author} {\bibfnamefont {G.~M.}\ \bibnamefont
			{Andolina}}, \bibinfo {author} {\bibfnamefont {A.}~\bibnamefont {Mari}},
		\bibinfo {author} {\bibfnamefont {M.}~\bibnamefont {Polini}},\ and\ \bibinfo
		{author} {\bibfnamefont {V.}~\bibnamefont {Giovannetti}},\ }\bibfield
	{title} {\bibinfo {title} {Charger-mediated energy transfer for quantum
			batteries: An open-system approach},\ }\href
	{https://doi.org/10.1103/PhysRevB.99.035421} {\bibfield  {journal} {\bibinfo
			{journal} {Phys. Rev. B}\ }\textbf {\bibinfo {volume} {99}},\ \bibinfo
		{pages} {035421} (\bibinfo {year} {2019})}\BibitemShut {NoStop}%
	\bibitem [{\citenamefont {Arjmandi}\ \emph
		{et~al.}(2022{\natexlab{a}})\citenamefont {Arjmandi}, \citenamefont {Shokri},
		\citenamefont {Faizi},\ and\ \citenamefont
		{Mohammadi}}]{arjmandi2022performance}%
	\BibitemOpen
	\bibfield  {author} {\bibinfo {author} {\bibfnamefont {M.~B.}\ \bibnamefont
			{Arjmandi}}, \bibinfo {author} {\bibfnamefont {A.}~\bibnamefont {Shokri}},
		\bibinfo {author} {\bibfnamefont {E.}~\bibnamefont {Faizi}},\ and\ \bibinfo
		{author} {\bibfnamefont {H.}~\bibnamefont {Mohammadi}},\ }\bibfield  {title}
	{\bibinfo {title} {Performance of quantum batteries with correlated and
			uncorrelated chargers},\ }\href {https://doi.org/10.1103/PhysRevA.106.062609}
	{\bibfield  {journal} {\bibinfo  {journal} {Phys. Rev. A}\ }\textbf {\bibinfo
			{volume} {106}},\ \bibinfo {pages} {062609} (\bibinfo {year}
		{2022}{\natexlab{a}})}\BibitemShut {NoStop}%
	\bibitem [{\citenamefont {Campaioli}\ \emph {et~al.}(2017)\citenamefont
		{Campaioli}, \citenamefont {Pollock}, \citenamefont {Binder}, \citenamefont
		{C\'eleri}, \citenamefont {Goold}, \citenamefont {Vinjanampathy},\ and\
		\citenamefont {Modi}}]{PRL2017Binder}%
	\BibitemOpen
	\bibfield  {author} {\bibinfo {author} {\bibfnamefont {F.}~\bibnamefont
			{Campaioli}}, \bibinfo {author} {\bibfnamefont {F.~A.}\ \bibnamefont
			{Pollock}}, \bibinfo {author} {\bibfnamefont {F.~C.}\ \bibnamefont {Binder}},
		\bibinfo {author} {\bibfnamefont {L.}~\bibnamefont {C\'eleri}}, \bibinfo
		{author} {\bibfnamefont {J.}~\bibnamefont {Goold}}, \bibinfo {author}
		{\bibfnamefont {S.}~\bibnamefont {Vinjanampathy}},\ and\ \bibinfo {author}
		{\bibfnamefont {K.}~\bibnamefont {Modi}},\ }\bibfield  {title} {\bibinfo
		{title} {Enhancing the charging power of quantum batteries},\ }\href
	{https://doi.org/10.1103/PhysRevLett.118.150601} {\bibfield  {journal}
		{\bibinfo  {journal} {Phys. Rev. Lett.}\ }\textbf {\bibinfo {volume} {118}},\
		\bibinfo {pages} {150601} (\bibinfo {year} {2017})}\BibitemShut {NoStop}%
	\bibitem [{\citenamefont {Binder}\ \emph {et~al.}(2015)\citenamefont {Binder},
		\citenamefont {Vinjanampathy}, \citenamefont {Modi},\ and\ \citenamefont
		{Goold}}]{Binder:15}%
	\BibitemOpen
	\bibfield  {author} {\bibinfo {author} {\bibfnamefont {F.~C.}\ \bibnamefont
			{Binder}}, \bibinfo {author} {\bibfnamefont {S.}~\bibnamefont
			{Vinjanampathy}}, \bibinfo {author} {\bibfnamefont {K.}~\bibnamefont
			{Modi}},\ and\ \bibinfo {author} {\bibfnamefont {J.}~\bibnamefont {Goold}},\
	}\bibfield  {title} {\bibinfo {title} {Quantacell: powerful charging of
			quantum batteries},\ }\href {https://doi.org/10.1088/1367-2630/17/7/075015}
	{\bibfield  {journal} {\bibinfo  {journal} {New J. Phys.}\ }\textbf {\bibinfo
			{volume} {17}},\ \bibinfo {pages} {075015} (\bibinfo {year}
		{2015})}\BibitemShut {NoStop}%
	\bibitem [{\citenamefont {Rossini}\ \emph {et~al.}(2020)\citenamefont
		{Rossini}, \citenamefont {Andolina}, \citenamefont {Rosa}, \citenamefont
		{Carrega},\ and\ \citenamefont {Polini}}]{Rossini:20}%
	\BibitemOpen
	\bibfield  {author} {\bibinfo {author} {\bibfnamefont {D.}~\bibnamefont
			{Rossini}}, \bibinfo {author} {\bibfnamefont {G.~M.}\ \bibnamefont
			{Andolina}}, \bibinfo {author} {\bibfnamefont {D.}~\bibnamefont {Rosa}},
		\bibinfo {author} {\bibfnamefont {M.}~\bibnamefont {Carrega}},\ and\ \bibinfo
		{author} {\bibfnamefont {M.}~\bibnamefont {Polini}},\ }\bibfield  {title}
	{\bibinfo {title} {Quantum advantage in the charging process of
			sachdev-ye-kitaev batteries},\ }\href
	{https://doi.org/10.1103/PhysRevLett.125.236402} {\bibfield  {journal}
		{\bibinfo  {journal} {Phys. Rev. Lett.}\ }\textbf {\bibinfo {volume} {125}},\
		\bibinfo {pages} {236402} (\bibinfo {year} {2020})}\BibitemShut {NoStop}%
	\bibitem [{\citenamefont {Crescente}\ \emph {et~al.}(2020)\citenamefont
		{Crescente}, \citenamefont {Carrega}, \citenamefont {Sassetti},\ and\
		\citenamefont {Ferraro}}]{Crescente:20}%
	\BibitemOpen
	\bibfield  {author} {\bibinfo {author} {\bibfnamefont {A.}~\bibnamefont
			{Crescente}}, \bibinfo {author} {\bibfnamefont {M.}~\bibnamefont {Carrega}},
		\bibinfo {author} {\bibfnamefont {M.}~\bibnamefont {Sassetti}},\ and\
		\bibinfo {author} {\bibfnamefont {D.}~\bibnamefont {Ferraro}},\ }\bibfield
	{title} {\bibinfo {title} {Ultrafast charging in a two-photon dicke quantum
			battery},\ }\href {https://doi.org/10.1103/PhysRevB.102.245407} {\bibfield
		{journal} {\bibinfo  {journal} {Phys. Rev. B}\ }\textbf {\bibinfo {volume}
			{102}},\ \bibinfo {pages} {245407} (\bibinfo {year} {2020})}\BibitemShut
	{NoStop}%
	\bibitem [{\citenamefont {Ferraro}\ \emph {et~al.}(2018)\citenamefont
		{Ferraro}, \citenamefont {Campisi}, \citenamefont {Andolina}, \citenamefont
		{Pellegrini},\ and\ \citenamefont {Polini}}]{Ferraro:18}%
	\BibitemOpen
	\bibfield  {author} {\bibinfo {author} {\bibfnamefont {D.}~\bibnamefont
			{Ferraro}}, \bibinfo {author} {\bibfnamefont {M.}~\bibnamefont {Campisi}},
		\bibinfo {author} {\bibfnamefont {G.~M.}\ \bibnamefont {Andolina}}, \bibinfo
		{author} {\bibfnamefont {V.}~\bibnamefont {Pellegrini}},\ and\ \bibinfo
		{author} {\bibfnamefont {M.}~\bibnamefont {Polini}},\ }\bibfield  {title}
	{\bibinfo {title} {High-power collective charging of a solid-state quantum
			battery},\ }\href {https://doi.org/10.1103/PhysRevLett.120.117702} {\bibfield
		{journal} {\bibinfo  {journal} {Phys. Rev. Lett.}\ }\textbf {\bibinfo
			{volume} {120}},\ \bibinfo {pages} {117702} (\bibinfo {year}
		{2018})}\BibitemShut {NoStop}%
	\bibitem [{\citenamefont {Francica}(2022)}]{Gianluca:22}%
	\BibitemOpen
	\bibfield  {author} {\bibinfo {author} {\bibfnamefont {G.}~\bibnamefont
			{Francica}},\ }\bibfield  {title} {\bibinfo {title} {Quantum correlations and
			ergotropy},\ }\href {https://doi.org/10.1103/PhysRevE.105.L052101} {\bibfield
		{journal} {\bibinfo  {journal} {Phys. Rev. E}\ }\textbf {\bibinfo {volume}
			{105}},\ \bibinfo {pages} {L052101} (\bibinfo {year} {2022})}\BibitemShut
	{NoStop}%
	\bibitem [{\citenamefont {Francica}\ \emph {et~al.}(2017)\citenamefont
		{Francica}, \citenamefont {Goold}, \citenamefont {Plastina},\ and\
		\citenamefont {Paternostro}}]{Gianluca:17}%
	\BibitemOpen
	\bibfield  {author} {\bibinfo {author} {\bibfnamefont {G.}~\bibnamefont
			{Francica}}, \bibinfo {author} {\bibfnamefont {J.}~\bibnamefont {Goold}},
		\bibinfo {author} {\bibfnamefont {F.}~\bibnamefont {Plastina}},\ and\
		\bibinfo {author} {\bibfnamefont {M.}~\bibnamefont {Paternostro}},\
	}\bibfield  {title} {\bibinfo {title} {Daemonic ergotropy: enhanced work
			extraction from quantum correlations},\ }\href
	{https://doi.org/10.1038/s41534-017-0012-8} {\bibfield  {journal} {\bibinfo
			{journal} {npj Quantum Information}\ }\textbf {\bibinfo {volume} {1}},\
		\bibinfo {pages} {12} (\bibinfo {year} {2017})}\BibitemShut {NoStop}%
	\bibitem [{\citenamefont {Quach}\ and\ \citenamefont {Munro}(2020)}]{James:20}%
	\BibitemOpen
	\bibfield  {author} {\bibinfo {author} {\bibfnamefont {J.~Q.}\ \bibnamefont
			{Quach}}\ and\ \bibinfo {author} {\bibfnamefont {W.~J.}\ \bibnamefont
			{Munro}},\ }\bibfield  {title} {\bibinfo {title} {Using dark states to charge
			and stabilize open quantum batteries},\ }\href
	{https://doi.org/10.1103/PhysRevApplied.14.024092} {\bibfield  {journal}
		{\bibinfo  {journal} {Phys. Rev. Applied}\ }\textbf {\bibinfo {volume}
			{14}},\ \bibinfo {pages} {024092} (\bibinfo {year} {2020})}\BibitemShut
	{NoStop}%
	\bibitem [{\citenamefont {Kamin}\ \emph {et~al.}(2020)\citenamefont {Kamin},
		\citenamefont {Tabesh}, \citenamefont {Salimi},\ and\ \citenamefont
		{Santos}}]{Kamin:20-2}%
	\BibitemOpen
	\bibfield  {author} {\bibinfo {author} {\bibfnamefont {F.~H.}\ \bibnamefont
			{Kamin}}, \bibinfo {author} {\bibfnamefont {F.~T.}\ \bibnamefont {Tabesh}},
		\bibinfo {author} {\bibfnamefont {S.}~\bibnamefont {Salimi}},\ and\ \bibinfo
		{author} {\bibfnamefont {A.~C.}\ \bibnamefont {Santos}},\ }\bibfield  {title}
	{\bibinfo {title} {Entanglement, coherence, and charging process of quantum
			batteries},\ }\href {https://doi.org/10.1103/PhysRevE.102.052109} {\bibfield
		{journal} {\bibinfo  {journal} {Phys. Rev. E}\ }\textbf {\bibinfo {volume}
			{102}},\ \bibinfo {pages} {052109} (\bibinfo {year} {2020})}\BibitemShut
	{NoStop}%
	\bibitem [{\citenamefont {Ghosh}\ \emph {et~al.}(2021)\citenamefont {Ghosh},
		\citenamefont {Chanda}, \citenamefont {Mal},\ and\ \citenamefont
		{Sen(De)}}]{Ghosh:21}%
	\BibitemOpen
	\bibfield  {author} {\bibinfo {author} {\bibfnamefont {S.}~\bibnamefont
			{Ghosh}}, \bibinfo {author} {\bibfnamefont {T.}~\bibnamefont {Chanda}},
		\bibinfo {author} {\bibfnamefont {S.}~\bibnamefont {Mal}},\ and\ \bibinfo
		{author} {\bibfnamefont {A.}~\bibnamefont {Sen(De)}},\ }\bibfield  {title}
	{\bibinfo {title} {Fast charging of a quantum battery assisted by noise},\
	}\href {https://doi.org/10.1103/PhysRevA.104.032207} {\bibfield  {journal}
		{\bibinfo  {journal} {Phys. Rev. A}\ }\textbf {\bibinfo {volume} {104}},\
		\bibinfo {pages} {032207} (\bibinfo {year} {2021})}\BibitemShut {NoStop}%
	\bibitem [{\citenamefont {Monsel}\ \emph {et~al.}(2020)\citenamefont {Monsel},
		\citenamefont {Fellous-Asiani}, \citenamefont {Huard},\ and\ \citenamefont
		{Auff\`eves}}]{Alexia:20}%
	\BibitemOpen
	\bibfield  {author} {\bibinfo {author} {\bibfnamefont {J.}~\bibnamefont
			{Monsel}}, \bibinfo {author} {\bibfnamefont {M.}~\bibnamefont
			{Fellous-Asiani}}, \bibinfo {author} {\bibfnamefont {B.}~\bibnamefont
			{Huard}},\ and\ \bibinfo {author} {\bibfnamefont {A.}~\bibnamefont
			{Auff\`eves}},\ }\bibfield  {title} {\bibinfo {title} {The energetic cost of
			work extraction},\ }\href {https://doi.org/10.1103/PhysRevLett.124.130601}
	{\bibfield  {journal} {\bibinfo  {journal} {Phys. Rev. Lett.}\ }\textbf
		{\bibinfo {volume} {124}},\ \bibinfo {pages} {130601} (\bibinfo {year}
		{2020})}\BibitemShut {NoStop}%
	\bibitem [{\citenamefont {Francica}\ \emph {et~al.}(2020)\citenamefont
		{Francica}, \citenamefont {Binder}, \citenamefont {Guarnieri}, \citenamefont
		{Mitchison}, \citenamefont {Goold},\ and\ \citenamefont
		{Plastina}}]{Francica:20}%
	\BibitemOpen
	\bibfield  {author} {\bibinfo {author} {\bibfnamefont {G.}~\bibnamefont
			{Francica}}, \bibinfo {author} {\bibfnamefont {F.~C.}\ \bibnamefont
			{Binder}}, \bibinfo {author} {\bibfnamefont {G.}~\bibnamefont {Guarnieri}},
		\bibinfo {author} {\bibfnamefont {M.~T.}\ \bibnamefont {Mitchison}}, \bibinfo
		{author} {\bibfnamefont {J.}~\bibnamefont {Goold}},\ and\ \bibinfo {author}
		{\bibfnamefont {F.}~\bibnamefont {Plastina}},\ }\bibfield  {title} {\bibinfo
		{title} {Quantum coherence and ergotropy},\ }\href
	{https://doi.org/10.1103/PhysRevLett.125.180603} {\bibfield  {journal}
		{\bibinfo  {journal} {Phys. Rev. Lett.}\ }\textbf {\bibinfo {volume} {125}},\
		\bibinfo {pages} {180603} (\bibinfo {year} {2020})}\BibitemShut {NoStop}%
	\bibitem [{\citenamefont {Santos}(2021)}]{Santos:21b}%
	\BibitemOpen
	\bibfield  {author} {\bibinfo {author} {\bibfnamefont {A.~C.}\ \bibnamefont
			{Santos}},\ }\bibfield  {title} {\bibinfo {title} {Quantum advantage of
			two-level batteries in the self-discharging process},\ }\href
	{https://doi.org/10.1103/PhysRevE.103.042118} {\bibfield  {journal} {\bibinfo
			{journal} {Phys. Rev. E}\ }\textbf {\bibinfo {volume} {103}},\ \bibinfo
		{pages} {042118} (\bibinfo {year} {2021})}\BibitemShut {NoStop}%
	\bibitem [{\citenamefont {{{\c{C}}akmak}}(2020)}]{Baris:20}%
	\BibitemOpen
	\bibfield  {author} {\bibinfo {author} {\bibfnamefont {B.}~\bibnamefont
			{{{\c{C}}akmak}}},\ }\bibfield  {title} {\bibinfo {title} {Ergotropy from
			coherences in an open quantum system},\ }\href
	{https://doi.org/10.1103/PhysRevE.102.042111} {\bibfield  {journal} {\bibinfo
			{journal} {Phys. Rev. E}\ }\textbf {\bibinfo {volume} {102}},\ \bibinfo
		{pages} {042111} (\bibinfo {year} {2020})}\BibitemShut {NoStop}%
	\bibitem [{\citenamefont {Shi}\ \emph {et~al.}(2022)\citenamefont {Shi},
		\citenamefont {Ding}, \citenamefont {Wan}, \citenamefont {Wang},\ and\
		\citenamefont {Yang}}]{shi2022entanglement}%
	\BibitemOpen
	\bibfield  {author} {\bibinfo {author} {\bibfnamefont {H.-L.}\ \bibnamefont
			{Shi}}, \bibinfo {author} {\bibfnamefont {S.}~\bibnamefont {Ding}}, \bibinfo
		{author} {\bibfnamefont {Q.-K.}\ \bibnamefont {Wan}}, \bibinfo {author}
		{\bibfnamefont {X.-H.}\ \bibnamefont {Wang}},\ and\ \bibinfo {author}
		{\bibfnamefont {W.-L.}\ \bibnamefont {Yang}},\ }\bibfield  {title} {\bibinfo
		{title} {Entanglement, coherence, and extractable work in quantum
			batteries},\ }\href {https://doi.org/10.1103/PhysRevLett.129.130602}
	{\bibfield  {journal} {\bibinfo  {journal} {Physical Review Letters}\
		}\textbf {\bibinfo {volume} {129}},\ \bibinfo {pages} {130602} (\bibinfo
		{year} {2022})}\BibitemShut {NoStop}%
	\bibitem [{\citenamefont {{Tirone}}\ \emph {et~al.}(2023)\citenamefont
		{{Tirone}}, \citenamefont {{Salvia}}, \citenamefont {{Chessa}},\ and\
		\citenamefont {{Giovannetti}}}]{2023arXiv230516803T}%
	\BibitemOpen
	\bibfield  {author} {\bibinfo {author} {\bibfnamefont {S.}~\bibnamefont
			{{Tirone}}}, \bibinfo {author} {\bibfnamefont {R.}~\bibnamefont {{Salvia}}},
		\bibinfo {author} {\bibfnamefont {S.}~\bibnamefont {{Chessa}}},\ and\
		\bibinfo {author} {\bibfnamefont {V.}~\bibnamefont {{Giovannetti}}},\
	}\bibfield  {title} {\bibinfo {title} {{Quantum work extraction efficiency
				for noisy quantum batteries: the role of coherence}},\ }\href
	{https://doi.org/10.48550/arXiv.2305.16803} {\bibfield  {journal} {\bibinfo
			{journal} {arXiv e-prints}\ ,\ \bibinfo {eid} {arXiv:2305.16803}} (\bibinfo
		{year} {2023})},\ \Eprint {https://arxiv.org/abs/2305.16803}
	{arXiv:2305.16803 [quant-ph]} \BibitemShut {NoStop}%
	\bibitem [{\citenamefont {Gyhm}\ \emph {et~al.}(2022)\citenamefont {Gyhm},
		\citenamefont {{\v{S}}afr{\'a}nek},\ and\ \citenamefont
		{Rosa}}]{gyhm2022quantum}%
	\BibitemOpen
	\bibfield  {author} {\bibinfo {author} {\bibfnamefont {J.-Y.}\ \bibnamefont
			{Gyhm}}, \bibinfo {author} {\bibfnamefont {D.}~\bibnamefont
			{{\v{S}}afr{\'a}nek}},\ and\ \bibinfo {author} {\bibfnamefont
			{D.}~\bibnamefont {Rosa}},\ }\bibfield  {title} {\bibinfo {title} {Quantum
			charging advantage cannot be extensive without global operations},\ }\href
	{https://doi.org/10.1103/PhysRevLett.128.140501} {\bibfield  {journal}
		{\bibinfo  {journal} {Physical Review Letters}\ }\textbf {\bibinfo {volume}
			{128}},\ \bibinfo {pages} {140501} (\bibinfo {year} {2022})}\BibitemShut
	{NoStop}%
	\bibitem [{\citenamefont {Joshi}\ and\ \citenamefont
		{Mahesh}(2022)}]{PhysRevA.106.042601}%
	\BibitemOpen
	\bibfield  {author} {\bibinfo {author} {\bibfnamefont {J.}~\bibnamefont
			{Joshi}}\ and\ \bibinfo {author} {\bibfnamefont {T.~S.}\ \bibnamefont
			{Mahesh}},\ }\bibfield  {title} {\bibinfo {title} {Experimental investigation
			of a quantum battery using star-topology nmr spin systems},\ }\href
	{https://doi.org/10.1103/PhysRevA.106.042601} {\bibfield  {journal} {\bibinfo
			{journal} {Phys. Rev. A}\ }\textbf {\bibinfo {volume} {106}},\ \bibinfo
		{pages} {042601} (\bibinfo {year} {2022})}\BibitemShut {NoStop}%
	\bibitem [{\citenamefont {Hu}\ \emph {et~al.}(2022)\citenamefont {Hu},
		\citenamefont {Qiu}, \citenamefont {Souza}, \citenamefont {Yuan},
		\citenamefont {Zhou}, \citenamefont {Zhang}, \citenamefont {Chu},
		\citenamefont {Pan}, \citenamefont {Hu}, \citenamefont {Li}, \citenamefont
		{Xu}, \citenamefont {Zhong}, \citenamefont {Liu}, \citenamefont {Yan},
		\citenamefont {Tan}, \citenamefont {Bachelard}, \citenamefont {Villas-Boas},
		\citenamefont {Santos},\ and\ \citenamefont {Yu}}]{Hu:21}%
	\BibitemOpen
	\bibfield  {author} {\bibinfo {author} {\bibfnamefont {C.-K.}\ \bibnamefont
			{Hu}}, \bibinfo {author} {\bibfnamefont {J.}~\bibnamefont {Qiu}}, \bibinfo
		{author} {\bibfnamefont {P.~J.~P.}\ \bibnamefont {Souza}}, \bibinfo {author}
		{\bibfnamefont {J.}~\bibnamefont {Yuan}}, \bibinfo {author} {\bibfnamefont
			{Y.}~\bibnamefont {Zhou}}, \bibinfo {author} {\bibfnamefont {L.}~\bibnamefont
			{Zhang}}, \bibinfo {author} {\bibfnamefont {J.}~\bibnamefont {Chu}}, \bibinfo
		{author} {\bibfnamefont {X.}~\bibnamefont {Pan}}, \bibinfo {author}
		{\bibfnamefont {L.}~\bibnamefont {Hu}}, \bibinfo {author} {\bibfnamefont
			{J.}~\bibnamefont {Li}}, \bibinfo {author} {\bibfnamefont {Y.}~\bibnamefont
			{Xu}}, \bibinfo {author} {\bibfnamefont {Y.}~\bibnamefont {Zhong}}, \bibinfo
		{author} {\bibfnamefont {S.}~\bibnamefont {Liu}}, \bibinfo {author}
		{\bibfnamefont {F.}~\bibnamefont {Yan}}, \bibinfo {author} {\bibfnamefont
			{D.}~\bibnamefont {Tan}}, \bibinfo {author} {\bibfnamefont {R.}~\bibnamefont
			{Bachelard}}, \bibinfo {author} {\bibfnamefont {C.~J.}\ \bibnamefont
			{Villas-Boas}}, \bibinfo {author} {\bibfnamefont {A.~C.}\ \bibnamefont
			{Santos}},\ and\ \bibinfo {author} {\bibfnamefont {D.}~\bibnamefont {Yu}},\
	}\bibfield  {title} {\bibinfo {title} {Optimal charging of a superconducting
			quantum battery},\ }\href {https://doi.org/10.1088/2058-9565/ac8444}
	{\bibfield  {journal} {\bibinfo  {journal} {Quantum Science and Technology}\
		}\textbf {\bibinfo {volume} {7}},\ \bibinfo {pages} {045018} (\bibinfo {year}
		{2022})}\BibitemShut {NoStop}%
	\bibitem [{\citenamefont {Gemme}\ \emph {et~al.}(2022)\citenamefont {Gemme},
		\citenamefont {Grossi}, \citenamefont {Ferraro}, \citenamefont {Vallecorsa},\
		and\ \citenamefont {Sassetti}}]{batteries8050043}%
	\BibitemOpen
	\bibfield  {author} {\bibinfo {author} {\bibfnamefont {G.}~\bibnamefont
			{Gemme}}, \bibinfo {author} {\bibfnamefont {M.}~\bibnamefont {Grossi}},
		\bibinfo {author} {\bibfnamefont {D.}~\bibnamefont {Ferraro}}, \bibinfo
		{author} {\bibfnamefont {S.}~\bibnamefont {Vallecorsa}},\ and\ \bibinfo
		{author} {\bibfnamefont {M.}~\bibnamefont {Sassetti}},\ }\bibfield  {title}
	{\bibinfo {title} {Ibm quantum platforms: A quantum battery perspective},\
	}\bibfield  {journal} {\bibinfo  {journal} {Batteries}\ }\textbf {\bibinfo
		{volume} {8}},\ \href {https://doi.org/10.3390/batteries8050043}
	{10.3390/batteries8050043} (\bibinfo {year} {2022})\BibitemShut {NoStop}%
	\bibitem [{\citenamefont {Cruz}\ \emph {et~al.}(2022)\citenamefont {Cruz},
		\citenamefont {Anka}, \citenamefont {Reis}, \citenamefont {Bachelard},\ and\
		\citenamefont {Santos}}]{Cruz:22}%
	\BibitemOpen
	\bibfield  {author} {\bibinfo {author} {\bibfnamefont {C.}~\bibnamefont
			{Cruz}}, \bibinfo {author} {\bibfnamefont {M.~F.}\ \bibnamefont {Anka}},
		\bibinfo {author} {\bibfnamefont {M.~S.}\ \bibnamefont {Reis}}, \bibinfo
		{author} {\bibfnamefont {R.}~\bibnamefont {Bachelard}},\ and\ \bibinfo
		{author} {\bibfnamefont {A.~C.}\ \bibnamefont {Santos}},\ }\bibfield  {title}
	{\bibinfo {title} {Quantum battery based on quantum discord at room
			temperature},\ }\href
	{https://doi.org/https://doi.org/10.1088/2058-9565/ac57f3} {\bibfield
		{journal} {\bibinfo  {journal} {Quantum Sci. Technol.}\ }\textbf {\bibinfo
			{volume} {7}},\ \bibinfo {pages} {025020} (\bibinfo {year}
		{2022})}\BibitemShut {NoStop}%
	\bibitem [{\citenamefont {Huang}\ \emph {et~al.}(2023)\citenamefont {Huang},
		\citenamefont {Wang}, \citenamefont {Xiao}, \citenamefont {Gao},
		\citenamefont {Lin},\ and\ \citenamefont {Xue}}]{PhysRevA.107.L030201}%
	\BibitemOpen
	\bibfield  {author} {\bibinfo {author} {\bibfnamefont {X.}~\bibnamefont
			{Huang}}, \bibinfo {author} {\bibfnamefont {K.}~\bibnamefont {Wang}},
		\bibinfo {author} {\bibfnamefont {L.}~\bibnamefont {Xiao}}, \bibinfo {author}
		{\bibfnamefont {L.}~\bibnamefont {Gao}}, \bibinfo {author} {\bibfnamefont
			{H.}~\bibnamefont {Lin}},\ and\ \bibinfo {author} {\bibfnamefont
			{P.}~\bibnamefont {Xue}},\ }\bibfield  {title} {\bibinfo {title}
		{Demonstration of the charging progress of quantum batteries},\ }\href
	{https://doi.org/10.1103/PhysRevA.107.L030201} {\bibfield  {journal}
		{\bibinfo  {journal} {Phys. Rev. A}\ }\textbf {\bibinfo {volume} {107}},\
		\bibinfo {pages} {L030201} (\bibinfo {year} {2023})}\BibitemShut {NoStop}%
	\bibitem [{\citenamefont {da~C~Filho}\ \emph {et~al.}(2022)\citenamefont
		{da~C~Filho}, \citenamefont {Izquierdo}, \citenamefont {Saguia},
		\citenamefont {Albash}, \citenamefont {Hen},\ and\ \citenamefont
		{Sarandy}}]{da2022localization}%
	\BibitemOpen
	\bibfield  {author} {\bibinfo {author} {\bibfnamefont {J.~L.}\ \bibnamefont
			{da~C~Filho}}, \bibinfo {author} {\bibfnamefont {Z.~G.}\ \bibnamefont
			{Izquierdo}}, \bibinfo {author} {\bibfnamefont {A.}~\bibnamefont {Saguia}},
		\bibinfo {author} {\bibfnamefont {T.}~\bibnamefont {Albash}}, \bibinfo
		{author} {\bibfnamefont {I.}~\bibnamefont {Hen}},\ and\ \bibinfo {author}
		{\bibfnamefont {M.~S.}\ \bibnamefont {Sarandy}},\ }\bibfield  {title}
	{\bibinfo {title} {Localization transition induced by programmable
			disorder},\ }\href {https://doi.org/10.1103/PhysRevB.105.134201} {\bibfield
		{journal} {\bibinfo  {journal} {Physical Review B}\ }\textbf {\bibinfo
			{volume} {105}},\ \bibinfo {pages} {134201} (\bibinfo {year}
		{2022})}\BibitemShut {NoStop}%
	\bibitem [{\citenamefont {Kj{\"a}ll}\ \emph {et~al.}(2014)\citenamefont
		{Kj{\"a}ll}, \citenamefont {Bardarson},\ and\ \citenamefont
		{Pollmann}}]{kjall2014many}%
	\BibitemOpen
	\bibfield  {author} {\bibinfo {author} {\bibfnamefont {J.~A.}\ \bibnamefont
			{Kj{\"a}ll}}, \bibinfo {author} {\bibfnamefont {J.~H.}\ \bibnamefont
			{Bardarson}},\ and\ \bibinfo {author} {\bibfnamefont {F.}~\bibnamefont
			{Pollmann}},\ }\bibfield  {title} {\bibinfo {title} {Many-body localization
			in a disordered quantum ising chain},\ }\href
	{https://doi.org/10.1103/PhysRevLett.113.107204} {\bibfield  {journal}
		{\bibinfo  {journal} {Physical review letters}\ }\textbf {\bibinfo {volume}
			{113}},\ \bibinfo {pages} {107204} (\bibinfo {year} {2014})}\BibitemShut
	{NoStop}%
	\bibitem [{\citenamefont {Ghosh}\ \emph {et~al.}(2020)\citenamefont {Ghosh},
		\citenamefont {Chanda}, \citenamefont {Sen} \emph
		{et~al.}}]{ghosh2020enhancement}%
	\BibitemOpen
	\bibfield  {author} {\bibinfo {author} {\bibfnamefont {S.}~\bibnamefont
			{Ghosh}}, \bibinfo {author} {\bibfnamefont {T.}~\bibnamefont {Chanda}},
		\bibinfo {author} {\bibfnamefont {A.}~\bibnamefont {Sen}}, \emph {et~al.},\
	}\bibfield  {title} {\bibinfo {title} {Enhancement in the performance of a
			quantum battery by ordered and disordered interactions},\ }\href
	{https://doi.org/10.1103/PhysRevA.101.032115} {\bibfield  {journal} {\bibinfo
			{journal} {Phys. Rev. A}\ }\textbf {\bibinfo {volume} {101}},\ \bibinfo
		{pages} {032115} (\bibinfo {year} {2020})}\BibitemShut {NoStop}%
	\bibitem [{\citenamefont {Rossini}\ \emph {et~al.}(2019)\citenamefont
		{Rossini}, \citenamefont {Andolina},\ and\ \citenamefont
		{Polini}}]{Andolina:19-2}%
	\BibitemOpen
	\bibfield  {author} {\bibinfo {author} {\bibfnamefont {D.}~\bibnamefont
			{Rossini}}, \bibinfo {author} {\bibfnamefont {G.~M.}\ \bibnamefont
			{Andolina}},\ and\ \bibinfo {author} {\bibfnamefont {M.}~\bibnamefont
			{Polini}},\ }\bibfield  {title} {\bibinfo {title} {Many-body localized
			quantum batteries},\ }\href {https://doi.org/10.1103/PhysRevB.100.115142}
	{\bibfield  {journal} {\bibinfo  {journal} {Phys. Rev. B}\ }\textbf {\bibinfo
			{volume} {100}},\ \bibinfo {pages} {115142} (\bibinfo {year}
		{2019})}\BibitemShut {NoStop}%
	\bibitem [{\citenamefont {Allahverdyan}\ \emph {et~al.}(2004)\citenamefont
		{Allahverdyan}, \citenamefont {Balian},\ and\ \citenamefont
		{Nieuwenhuizen}}]{Allahverdyan:04}%
	\BibitemOpen
	\bibfield  {author} {\bibinfo {author} {\bibfnamefont {A.~E.}\ \bibnamefont
			{Allahverdyan}}, \bibinfo {author} {\bibfnamefont {R.}~\bibnamefont
			{Balian}},\ and\ \bibinfo {author} {\bibfnamefont {T.~M.}\ \bibnamefont
			{Nieuwenhuizen}},\ }\bibfield  {title} {\bibinfo {title} {Maximal work
			extraction from finite quantum systems},\ }\href
	{https://doi.org/10.1209/epl/i2004-10101-2} {\bibfield  {journal} {\bibinfo
			{journal} {Europhys. Lett.}\ }\textbf {\bibinfo {volume} {67}},\ \bibinfo
		{pages} {565} (\bibinfo {year} {2004})}\BibitemShut {NoStop}%
	\bibitem [{\citenamefont {Moraes}\ \emph {et~al.}(2021)\citenamefont {Moraes},
		\citenamefont {Saguia}, \citenamefont {Santos},\ and\ \citenamefont
		{Sarandy}}]{Luiz:21}%
	\BibitemOpen
	\bibfield  {author} {\bibinfo {author} {\bibfnamefont {L.~F.~C.}\
			\bibnamefont {Moraes}}, \bibinfo {author} {\bibfnamefont {A.}~\bibnamefont
			{Saguia}}, \bibinfo {author} {\bibfnamefont {A.~C.}\ \bibnamefont {Santos}},\
		and\ \bibinfo {author} {\bibfnamefont {M.~S.}\ \bibnamefont {Sarandy}},\
	}\bibfield  {title} {\bibinfo {title} {Charging power and stability of
			always-on transitionless driven quantum batteries},\ }\href
	{http://iopscience.iop.org/article/10.1209/0295-5075/ac1363} {\bibfield
		{journal} {\bibinfo  {journal} {Accepted for publication in EPL (Europhysics
				Letters)}\ } (\bibinfo {year} {2021})}\BibitemShut {NoStop}%
	\bibitem [{\citenamefont {Arjmandi}\ \emph
		{et~al.}(2022{\natexlab{b}})\citenamefont {Arjmandi}, \citenamefont
		{Mohammadi},\ and\ \citenamefont {Santos}}]{arjmandi2022enhancing}%
	\BibitemOpen
	\bibfield  {author} {\bibinfo {author} {\bibfnamefont {M.~B.}\ \bibnamefont
			{Arjmandi}}, \bibinfo {author} {\bibfnamefont {H.}~\bibnamefont
			{Mohammadi}},\ and\ \bibinfo {author} {\bibfnamefont {A.~C.}\ \bibnamefont
			{Santos}},\ }\bibfield  {title} {\bibinfo {title} {Enhancing self-discharging
			process with disordered quantum batteries},\ }\href
	{https://doi.org/10.1103/PhysRevE.105.054115} {\bibfield  {journal} {\bibinfo
			{journal} {Phys. Rev. E}\ }\textbf {\bibinfo {volume} {105}},\ \bibinfo
		{pages} {054115} (\bibinfo {year} {2022}{\natexlab{b}})}\BibitemShut
	{NoStop}%
	\bibitem [{\citenamefont {Oliveira}\ \emph {et~al.}(2011)\citenamefont
		{Oliveira}, \citenamefont {Sarthour~Jr}, \citenamefont {Bonagamba},
		\citenamefont {Azevedo},\ and\ \citenamefont {Freitas}}]{Sarthour:Book}%
	\BibitemOpen
	\bibfield  {author} {\bibinfo {author} {\bibfnamefont {I.}~\bibnamefont
			{Oliveira}}, \bibinfo {author} {\bibfnamefont {R.}~\bibnamefont
			{Sarthour~Jr}}, \bibinfo {author} {\bibfnamefont {T.}~\bibnamefont
			{Bonagamba}}, \bibinfo {author} {\bibfnamefont {E.}~\bibnamefont {Azevedo}},\
		and\ \bibinfo {author} {\bibfnamefont {J.~C.}\ \bibnamefont {Freitas}},\
	}\href {https://doi.org/10.1016/B978-0-444-52782-0.X5000-3} {\emph {\bibinfo
			{title} {NMR quantum information processing}}}\ (\bibinfo  {publisher}
	{Elsevier},\ \bibinfo {address} {Oxford, UK},\ \bibinfo {year}
	{2011})\BibitemShut {NoStop}%
	\bibitem [{\citenamefont {{\v{Z}}nidari{\v{c}}}\ \emph
		{et~al.}(2008)\citenamefont {{\v{Z}}nidari{\v{c}}}, \citenamefont {Prosen},\
		and\ \citenamefont {Prelov{\v{s}}ek}}]{vznidarivc2008many}%
	\BibitemOpen
	\bibfield  {author} {\bibinfo {author} {\bibfnamefont {M.}~\bibnamefont
			{{\v{Z}}nidari{\v{c}}}}, \bibinfo {author} {\bibfnamefont {T.}~\bibnamefont
			{Prosen}},\ and\ \bibinfo {author} {\bibfnamefont {P.}~\bibnamefont
			{Prelov{\v{s}}ek}},\ }\bibfield  {title} {\bibinfo {title} {Many-body
			localization in the heisenberg x x z magnet in a random field},\ }\href
	{https://doi.org/10.1103/PhysRevB.77.064426} {\bibfield  {journal} {\bibinfo
			{journal} {Physical Review B}\ }\textbf {\bibinfo {volume} {77}},\ \bibinfo
		{pages} {064426} (\bibinfo {year} {2008})}\BibitemShut {NoStop}%
	\bibitem [{\citenamefont {Pal}\ and\ \citenamefont {Huse}(2010)}]{pal2010many}%
	\BibitemOpen
	\bibfield  {author} {\bibinfo {author} {\bibfnamefont {A.}~\bibnamefont
			{Pal}}\ and\ \bibinfo {author} {\bibfnamefont {D.~A.}\ \bibnamefont {Huse}},\
	}\bibfield  {title} {\bibinfo {title} {Many-body localization phase
			transition},\ }\href {https://doi.org/10.1103/PhysRevB.82.174411} {\bibfield
		{journal} {\bibinfo  {journal} {Physical review b}\ }\textbf {\bibinfo
			{volume} {82}},\ \bibinfo {pages} {174411} (\bibinfo {year}
		{2010})}\BibitemShut {NoStop}%
	\bibitem [{\citenamefont {Luitz}\ \emph {et~al.}(2015)\citenamefont {Luitz},
		\citenamefont {Laflorencie},\ and\ \citenamefont {Alet}}]{luitz2015many}%
	\BibitemOpen
	\bibfield  {author} {\bibinfo {author} {\bibfnamefont {D.~J.}\ \bibnamefont
			{Luitz}}, \bibinfo {author} {\bibfnamefont {N.}~\bibnamefont {Laflorencie}},\
		and\ \bibinfo {author} {\bibfnamefont {F.}~\bibnamefont {Alet}},\ }\bibfield
	{title} {\bibinfo {title} {Many-body localization edge in the random-field
			heisenberg chain},\ }\href {https://doi.org/10.1103/PhysRevB.91.081103}
	{\bibfield  {journal} {\bibinfo  {journal} {Physical Review B}\ }\textbf
		{\bibinfo {volume} {91}},\ \bibinfo {pages} {081103} (\bibinfo {year}
		{2015})}\BibitemShut {NoStop}%
	\bibitem [{\citenamefont {Hovhannisyan}\ \emph {et~al.}(2020)\citenamefont
		{Hovhannisyan}, \citenamefont {Barra},\ and\ \citenamefont
		{Imparato}}]{Hovhannisyan:20}%
	\BibitemOpen
	\bibfield  {author} {\bibinfo {author} {\bibfnamefont {K.~V.}\ \bibnamefont
			{Hovhannisyan}}, \bibinfo {author} {\bibfnamefont {F.}~\bibnamefont
			{Barra}},\ and\ \bibinfo {author} {\bibfnamefont {A.}~\bibnamefont
			{Imparato}},\ }\bibfield  {title} {\bibinfo {title} {Charging assisted by
			thermalization},\ }\href {https://doi.org/10.1103/PhysRevResearch.2.033413}
	{\bibfield  {journal} {\bibinfo  {journal} {Phys. Rev. Research}\ }\textbf
		{\bibinfo {volume} {2}},\ \bibinfo {pages} {033413} (\bibinfo {year}
		{2020})}\BibitemShut {NoStop}%
	\bibitem [{Note1()}]{Note1}%
	\BibitemOpen
	\bibinfo {note} {Since by using a Jordan-Wigner transformation~\cite
		{sachdev1999quantum} the model can be mapped into a non-interacting fermionic
		model, for which any arbitrary degree of disorder is able to localize the
		system~\cite {kjall2014many}}\BibitemShut {NoStop}%
	\bibitem [{\citenamefont {Santos}\ \emph
		{et~al.}(2020{\natexlab{a}})\citenamefont {Santos}, \citenamefont {Saguia},\
		and\ \citenamefont {Sarandy}}]{Santos:20c}%
	\BibitemOpen
	\bibfield  {author} {\bibinfo {author} {\bibfnamefont {A.~C.}\ \bibnamefont
			{Santos}}, \bibinfo {author} {\bibfnamefont {A.}~\bibnamefont {Saguia}},\
		and\ \bibinfo {author} {\bibfnamefont {M.~S.}\ \bibnamefont {Sarandy}},\
	}\bibfield  {title} {\bibinfo {title} {Stable and charge-switchable quantum
			batteries},\ }\href {https://doi.org/10.1103/PhysRevE.101.062114} {\bibfield
		{journal} {\bibinfo  {journal} {Phys. Rev. E}\ }\textbf {\bibinfo {volume}
			{101}},\ \bibinfo {pages} {062114} (\bibinfo {year}
		{2020}{\natexlab{a}})}\BibitemShut {NoStop}%
	\bibitem [{\citenamefont {Nandkishore}\ and\ \citenamefont
		{Huse}(2015)}]{nandkishore2015many}%
	\BibitemOpen
	\bibfield  {author} {\bibinfo {author} {\bibfnamefont {R.}~\bibnamefont
			{Nandkishore}}\ and\ \bibinfo {author} {\bibfnamefont {D.~A.}\ \bibnamefont
			{Huse}},\ }\bibfield  {title} {\bibinfo {title} {Many-body localization and
			thermalization in quantum statistical mechanics},\ }\href
	{https://doi.org/10.1146/annurev-conmatphys-031214-014726} {\bibfield
		{journal} {\bibinfo  {journal} {Annu. Rev. Condens. Matter Phys.}\ }\textbf
		{\bibinfo {volume} {6}},\ \bibinfo {pages} {15} (\bibinfo {year}
		{2015})}\BibitemShut {NoStop}%
	\bibitem [{\citenamefont {Krummheuer}\ \emph {et~al.}(2002)\citenamefont
		{Krummheuer}, \citenamefont {Axt},\ and\ \citenamefont
		{Kuhn}}]{krummheuer2002theory}%
	\BibitemOpen
	\bibfield  {author} {\bibinfo {author} {\bibfnamefont {B.}~\bibnamefont
			{Krummheuer}}, \bibinfo {author} {\bibfnamefont {V.~M.}\ \bibnamefont
			{Axt}},\ and\ \bibinfo {author} {\bibfnamefont {T.}~\bibnamefont {Kuhn}},\
	}\bibfield  {title} {\bibinfo {title} {Theory of pure dephasing and the
			resulting absorption line shape in semiconductor quantum dots},\ }\href
	{https://doi.org/10.1103/PhysRevB.65.195313} {\bibfield  {journal} {\bibinfo
			{journal} {Physical Review B}\ }\textbf {\bibinfo {volume} {65}},\ \bibinfo
		{pages} {195313} (\bibinfo {year} {2002})}\BibitemShut {NoStop}%
	\bibitem [{\citenamefont {Fan}\ \emph {et~al.}(1998)\citenamefont {Fan},
		\citenamefont {Takagahara}, \citenamefont {Cunningham},\ and\ \citenamefont
		{Wang}}]{fan1998pure}%
	\BibitemOpen
	\bibfield  {author} {\bibinfo {author} {\bibfnamefont {X.}~\bibnamefont
			{Fan}}, \bibinfo {author} {\bibfnamefont {T.}~\bibnamefont {Takagahara}},
		\bibinfo {author} {\bibfnamefont {J.}~\bibnamefont {Cunningham}},\ and\
		\bibinfo {author} {\bibfnamefont {H.}~\bibnamefont {Wang}},\ }\bibfield
	{title} {\bibinfo {title} {Pure dephasing induced by exciton--phonon
			interactions in narrow gaas quantum wells},\ }\href
	{https://doi.org/10.1016/S0038-1098(98)00461-X} {\bibfield  {journal}
		{\bibinfo  {journal} {Solid state communications}\ }\textbf {\bibinfo
			{volume} {108}},\ \bibinfo {pages} {857} (\bibinfo {year}
		{1998})}\BibitemShut {NoStop}%
	\bibitem [{\citenamefont {Besombes}\ \emph {et~al.}(2001)\citenamefont
		{Besombes}, \citenamefont {Kheng}, \citenamefont {Marsal},\ and\
		\citenamefont {Mariette}}]{besombes2001acoustic}%
	\BibitemOpen
	\bibfield  {author} {\bibinfo {author} {\bibfnamefont {L.}~\bibnamefont
			{Besombes}}, \bibinfo {author} {\bibfnamefont {K.}~\bibnamefont {Kheng}},
		\bibinfo {author} {\bibfnamefont {L.}~\bibnamefont {Marsal}},\ and\ \bibinfo
		{author} {\bibfnamefont {H.}~\bibnamefont {Mariette}},\ }\bibfield  {title}
	{\bibinfo {title} {Acoustic phonon broadening mechanism in single quantum dot
			emission},\ }\href {https://doi.org/10.1103/PhysRevB.63.155307} {\bibfield
		{journal} {\bibinfo  {journal} {Physical Review B}\ }\textbf {\bibinfo
			{volume} {63}},\ \bibinfo {pages} {155307} (\bibinfo {year}
		{2001})}\BibitemShut {NoStop}%
	\bibitem [{\citenamefont {Vagov}\ \emph {et~al.}(2004)\citenamefont {Vagov},
		\citenamefont {Axt}, \citenamefont {Kuhn}, \citenamefont {Langbein},
		\citenamefont {Borri},\ and\ \citenamefont
		{Woggon}}]{vagov2004nonmonotonous}%
	\BibitemOpen
	\bibfield  {author} {\bibinfo {author} {\bibfnamefont {A.}~\bibnamefont
			{Vagov}}, \bibinfo {author} {\bibfnamefont {V.~M.}\ \bibnamefont {Axt}},
		\bibinfo {author} {\bibfnamefont {T.}~\bibnamefont {Kuhn}}, \bibinfo {author}
		{\bibfnamefont {W.}~\bibnamefont {Langbein}}, \bibinfo {author}
		{\bibfnamefont {P.}~\bibnamefont {Borri}},\ and\ \bibinfo {author}
		{\bibfnamefont {U.}~\bibnamefont {Woggon}},\ }\bibfield  {title} {\bibinfo
		{title} {Nonmonotonous temperature dependence of the initial decoherence in
			quantum dots},\ }\href {https://doi.org/10.1103/PhysRevB.70.201305}
	{\bibfield  {journal} {\bibinfo  {journal} {Physical Review B}\ }\textbf
		{\bibinfo {volume} {70}},\ \bibinfo {pages} {201305} (\bibinfo {year}
		{2004})}\BibitemShut {NoStop}%
	\bibitem [{\citenamefont {Turchette}\ \emph {et~al.}(2000)\citenamefont
		{Turchette}, \citenamefont {Myatt}, \citenamefont {King}, \citenamefont
		{Sackett}, \citenamefont {Kielpinski}, \citenamefont {Itano}, \citenamefont
		{Monroe},\ and\ \citenamefont {Wineland}}]{turchette2000decoherence}%
	\BibitemOpen
	\bibfield  {author} {\bibinfo {author} {\bibfnamefont {Q.}~\bibnamefont
			{Turchette}}, \bibinfo {author} {\bibfnamefont {C.}~\bibnamefont {Myatt}},
		\bibinfo {author} {\bibfnamefont {B.}~\bibnamefont {King}}, \bibinfo {author}
		{\bibfnamefont {C.}~\bibnamefont {Sackett}}, \bibinfo {author} {\bibfnamefont
			{D.}~\bibnamefont {Kielpinski}}, \bibinfo {author} {\bibfnamefont
			{W.}~\bibnamefont {Itano}}, \bibinfo {author} {\bibfnamefont
			{C.}~\bibnamefont {Monroe}},\ and\ \bibinfo {author} {\bibfnamefont
			{D.}~\bibnamefont {Wineland}},\ }\bibfield  {title} {\bibinfo {title}
		{Decoherence and decay of motional quantum states of a trapped atom coupled
			to engineered reservoirs},\ }\href
	{https://doi.org/10.1103/PhysRevA.62.053807} {\bibfield  {journal} {\bibinfo
			{journal} {Physical Review A}\ }\textbf {\bibinfo {volume} {62}},\ \bibinfo
		{pages} {053807} (\bibinfo {year} {2000})}\BibitemShut {NoStop}%
	\bibitem [{\citenamefont {Myatt}\ \emph {et~al.}(2000)\citenamefont {Myatt},
		\citenamefont {King}, \citenamefont {Turchette}, \citenamefont {Sackett},
		\citenamefont {Kielpinski}, \citenamefont {Itano}, \citenamefont {Monroe},\
		and\ \citenamefont {Wineland}}]{myatt2000decoherence}%
	\BibitemOpen
	\bibfield  {author} {\bibinfo {author} {\bibfnamefont {C.~J.}\ \bibnamefont
			{Myatt}}, \bibinfo {author} {\bibfnamefont {B.~E.}\ \bibnamefont {King}},
		\bibinfo {author} {\bibfnamefont {Q.~A.}\ \bibnamefont {Turchette}}, \bibinfo
		{author} {\bibfnamefont {C.~A.}\ \bibnamefont {Sackett}}, \bibinfo {author}
		{\bibfnamefont {D.}~\bibnamefont {Kielpinski}}, \bibinfo {author}
		{\bibfnamefont {W.~M.}\ \bibnamefont {Itano}}, \bibinfo {author}
		{\bibfnamefont {C.}~\bibnamefont {Monroe}},\ and\ \bibinfo {author}
		{\bibfnamefont {D.~J.}\ \bibnamefont {Wineland}},\ }\bibfield  {title}
	{\bibinfo {title} {Decoherence of quantum superpositions through coupling to
			engineered reservoirs},\ }\href {https://doi.org/10.1038/35002001} {\bibfield
		{journal} {\bibinfo  {journal} {Nature}\ }\textbf {\bibinfo {volume}
			{403}},\ \bibinfo {pages} {269} (\bibinfo {year} {2000})}\BibitemShut
	{NoStop}%
	\bibitem [{\citenamefont {Santos}\ and\ \citenamefont
		{Sarandy}(2015)}]{Santos:15}%
	\BibitemOpen
	\bibfield  {author} {\bibinfo {author} {\bibfnamefont {A.~C.}\ \bibnamefont
			{Santos}}\ and\ \bibinfo {author} {\bibfnamefont {M.~S.}\ \bibnamefont
			{Sarandy}},\ }\bibfield  {title} {\bibinfo {title} {Superadiabatic controlled
			evolutions and universal quantum computation},\ }\href
	{https://doi.org/10.1038/srep15775} {\bibfield  {journal} {\bibinfo
			{journal} {Sci. Rep.}\ }\textbf {\bibinfo {volume} {5}},\ \bibinfo {pages}
		{15775} (\bibinfo {year} {2015})}\BibitemShut {NoStop}%
	\bibitem [{\citenamefont {Coulamy}\ \emph {et~al.}(2016)\citenamefont
		{Coulamy}, \citenamefont {Santos}, \citenamefont {Hen},\ and\ \citenamefont
		{Sarandy}}]{Coulamy:16}%
	\BibitemOpen
	\bibfield  {author} {\bibinfo {author} {\bibfnamefont {I.~B.}\ \bibnamefont
			{Coulamy}}, \bibinfo {author} {\bibfnamefont {A.~C.}\ \bibnamefont {Santos}},
		\bibinfo {author} {\bibfnamefont {I.}~\bibnamefont {Hen}},\ and\ \bibinfo
		{author} {\bibfnamefont {M.~S.}\ \bibnamefont {Sarandy}},\ }\bibfield
	{title} {\bibinfo {title} {Energetic cost of superadiabatic quantum
			computation},\ }\href {https://doi.org/10.3389/fict.2016.00019} {\bibfield
		{journal} {\bibinfo  {journal} {Frontiers in ICT}\ }\textbf {\bibinfo
			{volume} {3}},\ \bibinfo {pages} {19} (\bibinfo {year} {2016})}\BibitemShut
	{NoStop}%
	\bibitem [{\citenamefont {Campbell}\ and\ \citenamefont
		{Deffner}(2017)}]{Campbell-Deffner:17}%
	\BibitemOpen
	\bibfield  {author} {\bibinfo {author} {\bibfnamefont {S.}~\bibnamefont
			{Campbell}}\ and\ \bibinfo {author} {\bibfnamefont {S.}~\bibnamefont
			{Deffner}},\ }\bibfield  {title} {\bibinfo {title} {Trade-off between speed
			and cost in shortcuts to adiabaticity},\ }\href
	{https://doi.org/10.1103/PhysRevLett.118.100601} {\bibfield  {journal}
		{\bibinfo  {journal} {Phys. Rev. Lett.}\ }\textbf {\bibinfo {volume} {118}},\
		\bibinfo {pages} {100601} (\bibinfo {year} {2017})}\BibitemShut {NoStop}%
	\bibitem [{\citenamefont {Santos}\ and\ \citenamefont
		{Sarandy}(2018)}]{Santos:18-b}%
	\BibitemOpen
	\bibfield  {author} {\bibinfo {author} {\bibfnamefont {A.~C.}\ \bibnamefont
			{Santos}}\ and\ \bibinfo {author} {\bibfnamefont {M.~S.}\ \bibnamefont
			{Sarandy}},\ }\bibfield  {title} {\bibinfo {title} {Generalized shortcuts to
			adiabaticity and enhanced robustness against decoherence},\ }\href
	{https://doi.org/10.1088/1751-8121/aa96f1} {\bibfield  {journal} {\bibinfo
			{journal} {J. Phys. A: Math. Theor.}\ }\textbf {\bibinfo {volume} {51}},\
		\bibinfo {pages} {025301} (\bibinfo {year} {2018})}\BibitemShut {NoStop}%
	\bibitem [{\citenamefont {Hu}\ \emph {et~al.}(2018)\citenamefont {Hu},
		\citenamefont {Cui}, \citenamefont {Santos}, \citenamefont {Huang},
		\citenamefont {Sarandy}, \citenamefont {Li},\ and\ \citenamefont
		{Guo}}]{Hu:18}%
	\BibitemOpen
	\bibfield  {author} {\bibinfo {author} {\bibfnamefont {C.-K.}\ \bibnamefont
			{Hu}}, \bibinfo {author} {\bibfnamefont {J.-M.}\ \bibnamefont {Cui}},
		\bibinfo {author} {\bibfnamefont {A.~C.}\ \bibnamefont {Santos}}, \bibinfo
		{author} {\bibfnamefont {Y.-F.}\ \bibnamefont {Huang}}, \bibinfo {author}
		{\bibfnamefont {M.~S.}\ \bibnamefont {Sarandy}}, \bibinfo {author}
		{\bibfnamefont {C.-F.}\ \bibnamefont {Li}},\ and\ \bibinfo {author}
		{\bibfnamefont {G.-C.}\ \bibnamefont {Guo}},\ }\bibfield  {title} {\bibinfo
		{title} {Experimental implementation of generalized transitionless quantum
			driving},\ }\href {https://doi.org/10.1364/OL.43.003136} {\bibfield
		{journal} {\bibinfo  {journal} {Opt. Lett.}\ }\textbf {\bibinfo {volume}
			{43}},\ \bibinfo {pages} {3136} (\bibinfo {year} {2018})}\BibitemShut
	{NoStop}%
	\bibitem [{\citenamefont {Santos}\ \emph
		{et~al.}(2020{\natexlab{b}})\citenamefont {Santos}, \citenamefont {Nicotina},
		\citenamefont {Souza}, \citenamefont {Sarthour}, \citenamefont {Oliveira},\
		and\ \citenamefont {Sarandy}}]{Santos:20b}%
	\BibitemOpen
	\bibfield  {author} {\bibinfo {author} {\bibfnamefont {A.~C.}\ \bibnamefont
			{Santos}}, \bibinfo {author} {\bibfnamefont {A.}~\bibnamefont {Nicotina}},
		\bibinfo {author} {\bibfnamefont {A.~M.}\ \bibnamefont {Souza}}, \bibinfo
		{author} {\bibfnamefont {R.~S.}\ \bibnamefont {Sarthour}}, \bibinfo {author}
		{\bibfnamefont {I.~S.}\ \bibnamefont {Oliveira}},\ and\ \bibinfo {author}
		{\bibfnamefont {M.~S.}\ \bibnamefont {Sarandy}},\ }\bibfield  {title}
	{\bibinfo {title} {Optimizing {NMR} quantum information processing via
			generalized transitionless quantum driving},\ }\href
	{https://doi.org/10.1209/0295-5075/129/30008} {\bibfield  {journal} {\bibinfo
			{journal} {{EPL} (Europhysics Letters)}\ }\textbf {\bibinfo {volume}
			{129}},\ \bibinfo {pages} {30008} (\bibinfo {year}
		{2020}{\natexlab{b}})}\BibitemShut {NoStop}%
	\bibitem [{\citenamefont {Deffner}(2021)}]{Deffner:21}%
	\BibitemOpen
	\bibfield  {author} {\bibinfo {author} {\bibfnamefont {S.}~\bibnamefont
			{Deffner}},\ }\bibfield  {title} {\bibinfo {title} {Energetic cost of
			hamiltonian quantum gates},\ }\href
	{https://doi.org/10.1209/0295-5075/134/40002} {\bibfield  {journal} {\bibinfo
			{journal} {EPL (Europhysics Letters)}\ }\textbf {\bibinfo {volume} {134}},\
		\bibinfo {pages} {40002} (\bibinfo {year} {2021})}\BibitemShut {NoStop}%
	\bibitem [{\citenamefont {Sachdev}(1999)}]{sachdev1999quantum}%
	\BibitemOpen
	\bibfield  {author} {\bibinfo {author} {\bibfnamefont {S.}~\bibnamefont
			{Sachdev}},\ }\bibfield  {title} {\bibinfo {title} {Quantum phase
			transitions},\ }\href {https://doi.org/10.1088/2058-7058/12/4/23} {\bibfield
		{journal} {\bibinfo  {journal} {Physics world}\ }\textbf {\bibinfo {volume}
			{12}},\ \bibinfo {pages} {33} (\bibinfo {year} {1999})}\BibitemShut {NoStop}%
\end{thebibliography}

%

\end{document}